\documentclass[usenatbib]{mn2e}
\usepackage{graphicx,natbib,times,amsmath}
\usepackage{bm,url}
\graphicspath{{./fig/}{./png/}}
\usepackage{multirow}
\usepackage{color}
\usepackage{amssymb}

\newcommand{\EQ}{\begin{equation}}
\newcommand{\EN}{\end{equation}}
\newcommand{\EQA}{\begin{eqnarray}}
\newcommand{\ENA}{\end{eqnarray}}
\newcommand{\eq}[1]{(\ref{#1})}
\newcommand{\EEq}[1]{Equation~(\ref{#1})}
\newcommand{\Eq}[1]{Eq.~(\ref{#1})}
\newcommand{\Eqs}[2]{Eqs.~(\ref{#1}) and~(\ref{#2})}

\newcommand{\Sec}[1]{Sect.~\ref{#1}}

\newcommand{\Fig}[1]{Fig.~\ref{#1}}

\newcommand{\Tab}[1]{Table~\ref{#1}}

{}
{}
{}
\newcommand{\meanemf}{\overline{\cal E} {}}

{}
{}
\newcommand{\meanEMF}{\overline{\mbox{\boldmath ${\cal E}$}}{}}{}
\newcommand{\meanEEEE}{\overline{\mbox{\boldmath ${\cal E}$}}{}}{}
{}
{}
{}
{}
\newcommand{\meanAA}{\overline{\mbox{\boldmath $A$}}{}}{}
\newcommand{\meanBB}{\overline{\mbox{\boldmath $B$}}{}}{}
{}
{}
{}
{}
{}
{}
{}
{}
\newcommand{\meanJJ}{\overline{\mbox{\boldmath $J$}}{}}{}
{}
\newcommand{\meanUU}{\overline{\mbox{\boldmath $U$}}{}}{}

{}
{}
{}
\newcommand{\meanA}{\overline{A}}
\newcommand{\meanB}{\overline{B}}

\newcommand{\meanJ}{\overline{J}}

{}

{}
\newcommand{\emf}{{\cal E}}{}

\newcommand{\xx}{\bm{x}}

\newcommand{\bb}{\bm{b}}
\newcommand{\BB}{\bm{B}}

\newcommand{\uu}{\mbox{\boldmath $u$} {}}
\newcommand{\UU}{\mbox{\boldmath $U$} {}}

\newcommand{\ee}{\mbox{\boldmath $e$} {}}

\newcommand{\nab}{\mbox{\boldmath $\nabla$} {}}

\newcommand{\oo}{\mbox{\boldmath $\omega$} {}}

\newcommand{\ii}{{\rm i}}

\def\la{\mathrel{\mathchoice {\vcenter{\offinterlineskip\halign{\hfil
$\displaystyle##$\hfil\cr<\cr\sim\cr}}}
{\vcenter{\offinterlineskip\halign{\hfil$\textstyle##$\hfil\cr<\cr\sim\cr}}}
{\vcenter{\offinterlineskip\halign{\hfil$\scriptstyle##$\hfil\cr<\cr\sim\cr}}}
{\vcenter{\offinterlineskip\halign{\hfil$\scriptscriptstyle##$\hfil\cr<\cr\sim\cr}}}}}

\newcommand{\upim}{^{\rm i}}
\newcommand{\upre}{^{\rm r}}
\newcommand{\upresh}[1]{^{\raisebox{-#1pt}{$\,\scriptstyle {\rm r}$}}}
\newcommand{\upimsh}[1]{^{\raisebox{-#1pt}{$\,\scriptstyle {\rm i}$}}}
\def\Rm{R_{\rm m}}

\def\Rmc{R_{\rm m}^{\rm crit}}

\def\Brms{B_{\rm rms}}

\def\urms{u_{\rm rms}}

\def\brms{b_{\rm rms}}

\def\etat{\eta_{\rm t}}
\def\etath{\hat{\eta}_{\rm t}}
\def\etatz{\eta_{\rm t0}}
\def\etatz{\eta_{\rm t0}}
\def\etaTz{\eta_{\rm T0}}

\def\half{{\textstyle{1\over2}}}

\newcommand{\ymonber}[3]{ #1, {Monats.\ Dt.\ Akad.\ Wiss.,} {#2}, #3}
\newcommand{\yapj}[3]{ #1, {ApJ,} {#2}, #3}

\newcommand{\yan}[3]{ #1, {Astron.\ Nachr.,} {#2}, #3}

\newcommand{\yana}[3]{ #1, {A\&A,} {#2}, #3}

\newcommand{\ygafd}[3]{ #1, {Geophys.\ Astrophys.\ Fluid Dyn.,} {#2}, #3}

\newcommand{\ypf}[3]{ #1, {Phys.\ Fluids,} {#2}, #3}

\newcommand{\yprl}[3]{ #1, {Phys.\ Rev.\ Lett.,} {#2}, #3}

\newcommand{\yptrs}[3]{ #1, {Phil.\ Trans.\ R.\ Soc.,} {#2}, #3}

\newcommand{\ymn}[3]{ #1, {MNRAS,} {#2}, #3}

\newcommand{\ypre}[3]{ #1, {Phys.\ Rev.\ E,} {#2}, #3}

\newcommand{\yjour}[4]{ #1, {#2}, {#3}, #4}

\newcommand{\ybook}[3]{ #1, {#2} (#3)}
\newcommand{\yproc}[5]{ #1, in {#3}, ed.\ #4 (#5), #2}

\title[]{Mean-field dynamo action from delayed transport}
\author[M.\ Rheinhardt et al.]
{Matthias Rheinhardt$^1$\thanks{E-mail:mreinhardt@nordita.org},
Ebru Devlen$^{1,2}$, Karl-Heinz R\"adler$^{1,3}$, and
Axel Brandenburg$^{1,4}$
\\
$^1$Nordita, KTH Royal Institute of Technology and Stockholm University,
Roslagstullsbacken 23, SE-10691 Stockholm, Sweden\\
$^2$Department of Astronomy \& Space Sciences, Faculty of Science,
University of Ege, Bornova 35100, Izmir, Turkey\\
$^3$Leibniz-Institut f\"ur Astrophysik Potsdam, An der Sternwarte 16,
D-14482 Potsdam, Germany\\
$^4$Department of Astronomy, Stockholm University, SE-10691 Stockholm, Sweden
}

\date{\today,~ $ $Revision: 1.132 $ $}
\begin{document}
\maketitle

\label{firstpage}

\begin{abstract}
We analyze the nature of dynamo action that produces horizontally averaged
magnetic fields in two particular flows that were studied by Roberts
(1972, Phil.\ Trans.\ R.\ Soc.\ A 271, 411), namely his flows II and III.
They have zero kinetic helicity either pointwise (flow II),
or on average (flow III).
Using direct numerical simulations, we determine the onset conditions for
dynamo action at moderate values of the magnetic Reynolds number.
Using the test-field method, we show that the turbulent
magnetic diffusivity is then positive for both flows.
However, we demonstrate that for both flows large-scale dynamo action
occurs through delayed transport.
Mathematically speaking, the magnetic field at earlier times
contributes to the electromotive force through
the off-diagonal components of the $\alpha$ tensor
such that a zero mean magnetic field becomes unstable
to dynamo action.
This represents a qualitatively new mean-field dynamo mechanism not
previously described.
\end{abstract}


\section{Introduction}

The magnetic fields in various astrophysical settings are generally
believed to be produced by dynamo processes, which convert
kinetic energy into magnetic.
Small-scale dynamos produce magnetic energy at scales smaller
than or equal to that of the underlying motions,
large-scale dynamos at larger scales.
Both types of dynamos play important roles in astrophysics.
We may characterize large-scale dynamos
by the governing mechanism in the corresponding mean-field description.
One of the best known of these
{\it mean-field effects} is the $\alpha$ effect.
It quantifies the component of the mean electromotive force along
the direction of the mean magnetic field \citep{Par55,SKR66}, which
can lead to self-excitation.
In the presence of shear, the $\alpha$ effect can give rise
to traveling waves -- relevant to explaining the solar butterfly diagram.
Another important effect is turbulent diffusion,
described by the turbulent diffusivity $\etat$,
which quantifies a contribution to the mean electromotive force along the direction of the mean
current density.
In the absence of shear it is the balance of $\alpha$ effect vs.\ turbulent and microphysical
diffusion that determines the onset of dynamo action and, for oscillatory magnetic fields, also their period.
However, this basic picture of astrophysical large-scale dynamos is a strong simplification.
Both $\alpha$ effect and turbulent diffusivity are in general described by tensors.
This aspect is often ignored, in particular because
$\alpha$ effect dynamos work already under simple conditions,
under which these tensor properties are less important.

Dynamos based on the $\alpha$ effect are not the only ones.
Well-known alternatives include the $\Omega\times J$ effect \citep{Rae69,Rae69b,KR80}
and the shear--current effect \citep{RK03,RK04}, which rely upon
the existence of certain off-diagonal components of the $\etat$ tensor.
Another class of large-scale dynamos whose operation is
based upon the turbulent diffusivity tensor alone,
is due to negative turbulent diffusivity \citep{LNVW99,ZPF01,Zhe12}:
$\etat$ does not only become negative, but can even overcompensate the (positive) microphysical diffusivity.
Such dynamos have been studied using asymptotic analysis and have
only recently been confirmed in direct numerical simulations \citep{DBM13}.
A simple example of a flow capable of dynamo action of this type
is known as the Roberts-IV flow, which is one
of the flows studied in the seminal paper of \cite{Rob72}.
However, a proper description of such dynamos in terms of mean-field theory
is not straightforward because a negative total diffusivity
would destabilize modes on all scales with growth rates diverging with increasing wavenumber.
Luckily, in the case of the Roberts-IV flow it turned out that the turbulent diffusivity is effectively
wavenumber-dependent and negative only at small wavenumbers \citep{DBM13}.
Negative diffusivity dynamos are remarkable in the sense that the evolution
of different components of the mean magnetic field decouples.
This is not the case for $\alpha$ effect dynamos, nor those based on the $\Omega\times J$ and shear--current effects,
for which the mutual interaction between two components of the mean field is essential.
By contrast, in a negative diffusivity dynamo, one component can grow with the other permanently vanishing.

The dynamos mentioned so far are mean-field dynamos operating via an
instantaneous connection between the mean electromotive force $\meanEMF$
and the magnetic field $\meanBB$ or its (first) spatial derivatives.
We know, however, that an instantaneous connection is only an idealization \citep{Rae76} and that turbulent transport
has in general a memory effect, i.e., the electromotive force depends through a convolution on the values of the
mean magnetic field at all past times \citep{HB09}.
Although in isotropic turbulence such effects have been found to be small \citep{HB09},
examples have been given where they can be important \citep{HB09,Rae11,DBM13}.

In his seminal paper, \cite{Rob72} studied four simple spatially periodic steady flows in view of their dynamo action.
Flow I gives an often mentioned example for the classical $\alpha$ effect (e.g., \cite{Rae02} or \cite{Rae03}).
As \cite{DBM13} recognized, flow IV constitutes, if considered on the mean-field level, the above-mentioned dynamo
due to negative magnetic eddy diffusivity.
In the present paper we will analyze the dynamo mechanisms in flows II and III,
using direct numerical simulations (DNS) combined with analytic calculations
in the second order correlation approximation (SOCA) and the test-field method (TFM)
to compute the relevant transport coefficients.

In section 2 we define the flows, introduce the mean-field concept
and analyze their dynamo-relevant properties under SOCA.
In section 3 we first present numerical findings on dynamo action in flows II and III and then
provide explanations in mean-field terms relying upon the results of the TFM. Section 4 is devoted to the development
of a dynamical equation for the mean electromotive force occurring with flow II, while we draw conclusions in section 5.

\section{The problem considered}

\subsection{The Roberts flows}

\cite{Rob72} investigated four incompressible
spatially periodic steady flows
with regard to their dynamo properties.
More precisely, the flows vary periodically in the $x$ and $y$ directions,
but are independent of $z$.
We may write the corresponding velocities $\uu$ so that the components $u_x$ and $u_y$ have in all four cases the form
\EQ
u_x = v_0 \, \sin k_0x \, \cos k_0y \, , \quad u_y = - v_0 \, \cos k_0x \, \sin k_0y \, ,
\label{KH21}
\EN
while the components $u_z$ are different and given by
\begin{alignat}{3}
u_z\!&=&&\;w_0 \, \sin k_0x \, \sin k_0y
\quad&&\mbox{(flow I)},
\label{KH23}
\\
u_z\!&=&&\;w_0 \, \cos k_0x \, \cos k_0y
\quad&&\mbox{(flow II)},
\label{KH25}
\\
u_z\!&=&&\;\half w_0 (\cos 2 k_0x + \cos 2 k_0y)
\quad&&\mbox{(flow III)},
\label{KH27}
\\
u_z\!&=&&\;w_0 \, \sin k_0x
\quad&&\mbox{(flow IV)},
\label{KH29}
\end{alignat}
where $v_0$, $w_0$ and $k_0$ are constants.
In all four cases, Roberts found conditions under which dynamo action
is possible, that is, magnetic fields may grow.
The resulting magnetic fields survive $xy$ averaging and
are therefore amenable to mean-field treatment.

In view of mean-field dynamo theory, it is informative to consider
the kinetic helicity density
$h=\uu \cdot (\nab \times \uu)$ of the flows.
In the case of flow I, the volume average of $h$ is equal to $v_0 w_0 k_0$,
that is, in general non-zero.
As discussed in various contexts, we have then
an $\alpha$ effect \citep[e.g.][]{Rae02,Rae03},
which enables self-excitation of mean magnetic fields being of Beltrami type
by a so-called $\alpha^2$ dynamo.
Remarkably, in the case of flow II, $h$
vanishes everywhere (not only on average).
Nevertheless, as we will see, some kind of $\alpha$ effect occurs, which
explains the existence of mean-field dynamos, showing, however,
independent growth of its field components.
In flows III and IV, the mean kinetic helicity density $\overline{h}$
vanishes and we may not have an $\alpha$ effect.
While mean field dynamo action from flow IV has been demonstrated
as being due to negative magnetic eddy diffusivity \citep{DBM13},
we will show in this paper that flow III gives rise to self-excitation of mean fields
as a consequence of turbulent pumping, i.e., a $\gamma$ effect.
As for flow II, the field components evolve  independently.

\subsection{Mean-field modelling}

We consider the behavior of a magnetic field $\BB$ in an
infinitely extended homogeneous electrically conducting fluid
moving with a velocity $\UU$.
Then $\BB$ is governed by the induction equation
\EQ
\eta \nab^2 \BB + \nab \times (\UU \times \BB) - \partial_t \BB = {\bf 0} \, ,  \quad \nab \cdot \BB = 0 \, ,
\label{KH01}
\EN
where $\eta$ is the magnetic diffusivity of the fluid.

We adopt the concept of mean-field electrodynamics,
define mean fields as averages over all $x$ and $y$,
denote them by overbars, e.g., $\meanBB$ and $\meanUU$,
and put $\BB = \meanBB + \bb$ and $\UU = \meanUU + \uu$.
Clearly, mean fields like
$\meanBB$ and $\meanUU$ may then depend on $z$ and $t$ only.
We exclude here, however, a mean flow of the fluid, i.e. $\meanUU = {\bf 0}$,
and specify $\uu$ to be one of the flows introduced above.

From the induction equation \eq{KH01} we may derive its mean-field version
\EQ
\eta \nab^2 \meanBB + \nab \times \meanEMF - \partial_t \meanBB = {\bf 0} \, ,  \quad \nab \cdot \meanBB = 0 \, ,
\label{KH03}
\EN
with the mean electromotive force $\meanEMF$ defined by
\EQ
\meanEMF = \overline{\uu \times \bb} \, .
\label{KH05}
\EN
From \eq{KH01} and \eq{KH03} we may conclude that $\bb$ has to obey
\EQ
\eta \nab^2 \bb + \nab \times (\uu \times \bb)' - \partial_t \bb = - \nab \times (\uu \times \meanBB) \, ,
    \quad \nab \cdot \bb = 0 \, ,
\label{KH07}
\EN
where $(\uu \times \bb)'$ stands for $\uu \times \bb - \overline{\uu \times \bb}$.

If $\uu$ is specified according to \eq{KH21} and one of the relations \eq{KH23}--\eq{KH29},
$\bb$ and therefore $\meanEMF$ depend on the magnetic Reynolds numbers
$v_0 / \eta k_0$ and $w_0 / \eta k_0$.
For simplicity we define only one magnetic Reynolds number, $\Rm$, by
\EQ
\Rm = \max(v_0, w_0) / \eta k_0 \, .
\label{KH08}
\EN
Note the difference to the more common definition employing $\urms$.

Given that \eqref{KH07} is linear, $\bb$ is a linear functional of $\meanBB$ and its derivatives.
Under our assumptions, all spatial derivatives of $\meanBB$ can be expressed by the mean electric current density
$\meanJJ = (1/\mu) \nab \times \meanBB$, where $\mu$ means the magnetic permeability,
and we have simply $\meanJ_j = (1/\mu) \epsilon_{j3l} \partial_z \meanB_l$, that is,
$\meanJJ = (1/\mu) (-\partial_z \meanB_y, \partial_z \meanB_x, 0)$.
Hence for $\uu$ independent  of $z$, the mean electromotive force $\meanEMF$ can be represented in the form
\begin{align}
\meanemf_i (z,t) = \int \!\! \int &\big( a_{ij} (\zeta, \tau) \meanB_j (z-\zeta, t-\tau)
\nonumber\\
& - \eta_{ij} (\zeta, \tau) \mu \meanJ_j (z-\zeta, t-\tau) \big) \, d \zeta \, d \tau \, .
\label{KH11}
\end{align}
Here $a_{ij}$ and $\eta_{ij}$ are tensors, which are
symmetric\footnote{In order to see this, start from the more general relation\\
$\meanemf_i = \int \int K_{ij} (\zeta, \tau) \meanB_j (z - \zeta, t -\tau) \, d \zeta \, d \tau$,
split $K_{ij}$ into a part $a_{ij}$ that is symmetric and another one which is antisymmetric in $\zeta$.
Represent the latter one as a derivative of a quantity symmetric in $\zeta$.
An integration by parts delivers then a term of the type $b_{ij} \partial_z \meanB_j$ in the integrand,
with $b_{ij}$ being symmetric in $\zeta$.
It can easily be rewritten so that it takes the form of the $\eta_{ij} \mu \meanJ_j$ term in the integrand of \eq{KH11},
with $\eta_{ij}$ being symmetric in $\zeta$.}
in $\zeta$.

We may subject equations like \eq{KH03} and \eq{KH11} to a Fourier transformation with respect to $z$ and $t$,
\EQ
F (z, t) = \int \!\! \int \hat{F} (k, \omega) \exp\big(\ii (k z - \omega t)\big) \, d k \, d \omega \, .
\label{KH13}
\EN
Then \eq{KH03} turn into
\EQ
(\eta k^2 - \ii \omega)\hat{\meanBB} - \ii k \ee \times \hat{\meanEMF} = {\bm0} \, , \quad \hat{\meanB}_z = 0 \, ,
\label{KH14}
\EN
with \ee being the unit vector in the $z$ direction.
Here, of course, only the $x$ and $y$ components
of the first equation are of interest.
Equation \eq{KH11} turns into
\EQ
\hat{\meanemf}_i (k, \omega) = \hat{a}_{ij} (k, \omega) \hat{\meanB}_j (k, \omega)
    - \hat{\eta}_{ij} (k, \omega) \mu \hat{\meanJ}_j (k, \omega) \, .
\label{KH15}
\EN
The aforementioned symmetry of $a_{ij}$ and $\eta_{ij}$ in $\zeta$
occurs now as symmetry of $\hat{a}_{ij}$ and $\hat{\eta}_{ij}$ in $k$.
We restrict therefore all discussions about these and related quantities
to $k \geq 0$.
The imaginary parts of $\hat{a}_{ij}$ and $\hat{\eta}_{ij}$ vanish at $\omega = 0$.
Further we have $\hat{\meanJ}_j = (\ii k / \mu) \epsilon_{j3l} \hat{\meanB}_l$,
that is,
$\hat{\meanJJ} =  (\ii k / \mu) (-\hat{\meanB}_y, \hat{\meanB}_x, 0)$.

When using the Fourier transformation, we have to exclude functions
that grow exponentially in time.
If such functions occur, we may easily modify our considerations by
using a Laplace transformation instead;
see \cite{HB09} for examples.
Then $- \ii \omega$ is replaced by a complex variable, say $s$.

\subsection{Second-order correlation approximation}

In what follows we will sometimes refer to the {\it second-order correlation approximation} (SOCA),
which is defined by omitting the term with $(\uu \times \bb)'$ in \eq{KH07}.
As long as $\meanBB$ is steady or does not vary markedly during the time $(v_0 k_0)^{-1}$,
a sufficient condition for the applicability of this approximation reads $\Rm \ll 1$.
If $\meanBB$ varies more rapidly, this condition has to be replaced by $\max(v_0,w_0) k_0 \tau_0 \ll 1$,
where $\tau_0$ is a characteristic time of this variation.

For the determination of $\hat{\meanEMF}$ under SOCA,
we may use the Fourier-transformed versions of relations \eq{KH05} and \eq{KH07},
simplified by omitting the term $(\uu \times \bb)'$, that is,
\begin{align}
&\hat{\meanEMF} =\overline{\uu \times \hat{\bb}} \, , \label{KH31} \\
&\big( \eta (\partial_x^2  +\partial_y^2 - k^2) + \ii \omega \big) \hat{\bb}
= - (\hat{\meanB}_x \partial_x  + \hat{\meanB}_y \partial_y) \uu + \ii k \, u_z \hat{\meanBB} \, . \nonumber
\end{align}
We recall here that $\hat{\meanB}_z = 0$.

A straightforward calculation on the basis of \eq{KH31} with $\uu$ specified as flow I
leads to the relation \eq{KH15} with
\EQ
\begin{alignedat}{3}
 \hat{a}_{11} &= \hat{a}_{22}&&= \hat{\alpha} \, , \quad \hat{\alpha} &&= \frac{v_0 w_0 k_0}{2 \big(\eta (2 k_0^2 + k^2) - \ii \omega\big)} \, ,\\
 \hat{\eta}_{11} &= \hat{\eta}_{22} &&= \hat{\eta}_{\rm t} \, ,
    \quad \hat{\eta}_{\rm t} &&= \frac{w_0^2}{4 \big(\eta (2 k_0^2 + k^2) - \ii \omega\big)} \, .
\end{alignedat}
\label{KH33}
\EN
All other components of $\hat{a}_{ij}$ and $\hat{\eta}_{ij}$ are equal to zero.
The corresponding result for flow II differs from that only in so far as
$\hat{a}_{11}$ and $\hat{a}_{22}$ now vanish and the first relation
of \eq{KH33} has to be replaced by
\EQ
\hat{a}_{12} = \hat{a}_{21} = \hat{\alpha} \, .
\label{KH35}
\EN
All other relations \eq{KH33} remain valid and so also the remark that all not explicitly mentioned components
of $\hat{a}_{ij}$ and $\hat{\eta}_{ij}$ are equal to zero.

As for flows III and IV, all components of $\hat{a}_{ij}$ vanish and again also all of $\hat{\eta}_{ij}$,
except $\hat{\eta}_{11}$ and $\hat{\eta}_{22}$.
Putting
\EQ
\hat{\eta}_{11} = \hat{\eta}_{22} = \hat{\eta}_{\rm t} \, ,
\label{KH37}
\EN
we now have for flow III
\EQ
\hat{\eta}_{\rm t} = \frac{w_0^2}{4 \big(\eta (4 k_0^2 + k^2) - \ii \omega\big)}
\label{KH39}
\EN
and for flow IV
\EQ
\hat{\eta}_{\rm t} = \frac{w_0^2}{2 \big(\eta (k_0^2 + k^2) - \ii \omega\big)} \, .
\label{KH41}
\EN

We conclude from these results that, as long as SOCA applies, in the case of flow I
we have coupled equations for $\meanB_x$ and $\meanB_y$.
For flows II--IV, however, the equations for $\meanB_x$ and $\meanB_y$ are decoupled,
that is, $\meanB_x$ and $\meanB_y$ develop independently of each other.
The contributions $\ii \omega$ to the denominators in \eq{KH33}, \eq{KH39} and \eq{KH41} indicate
that memory effects occur,
that is, $\meanEMF$ at a given time depends also on $\meanBB$ at former times;
see \cite{HB09}, and in particular their Appendix~A.

Inserting our results for $\hat{\meanEMF}$ into the equations \eq{KH14} governing $\hat{\meanBB}$,
dispersion relations can be obtained.
Changing from Fourier to Laplace transformation with respect to $t$,
we replace $-\ii \omega$ by a complex variable $p$
so that a positive real part of $p$ means a growing solution.
In the case of flow I the dispersion relation reads
\EQ
p = \pm k \hat{\alpha} - (\eta + \hat{\eta}_{\rm t}) k^2 \, .
\label{KH51}
\EN
In the case of flow II, we have
\EQ
p = \mp \ii k\hat{\alpha} - (\eta + \hat{\eta}_{\rm t}) k^2 \, .
\label{KH53}
\EN
In the latter case, the upper and lower signs apply
for $\hat{\meanB}_x$ and $\hat{\meanB}_y$, respectively.
In the case of flows III and IV, \eqref{KH53} applies with $\hat{\alpha} = 0$.

The above dispersion relation \eq{KH51} for flow I combined with \eq{KH33},
allows steady or monotonously growing magnetic fields
for arbitrarily small $\Rm$ if only $k/k_0$ is sufficiently small.
(Decaying solutions can also be oscillatory.)  
In the case of flow II, we may conclude from \eq{KH53} and
\eq{KH33}, modified by \eq{KH35},
that the smallest value of $\Rm$ that allows growing magnetic fields
is obtained for $k/k_0 \to 0$.
For a marginally stable field and $v_0=w_0$ we have in this limit $\Rm = 2 \sqrt{2}$.
This field is oscillating with a frequency $\omega = 2 \eta k_0 k$.
With this value of $\Rm$, however, we are beyond the validity range of SOCA.
In the case of flows III and IV, we find no solutions of the above dispersion relations,
that is, \eq{KH53} with $\hat{\alpha} = 0$ and \eq{KH39} or \eq{KH41},
that would correspond to marginally stable or growing magnetic fields even if SOCA were valid.

\subsection{Possibility of a dynamo from time delay}

Let us consider a simple example which shows how the memory effect makes a dynamo possible.
Assume, thinking of flow II, that a component of $\meanBB$, say $\meanB_x$, is independent of the others,
depends only on $z$ and $t$, and obeys
\EQ
\eta \partial^2_z \meanB_x - \partial_z \meanemf_y - \partial_t \meanB_x = 0 \, .
\label{example}
\EN
Ignoring first the memory effect, we put $\meanemf_y = \alpha \meanB_x$
with $\alpha$ independent of $z$ and $t$, but ignore for simplicity $\etat$.
Without loss of generality we may restrict ourselves to solutions $\meanB_x$ of \eq{example}
that are proportional to  $\exp (\ii k z + p t)$.
We have then $\mbox{Re}\,p = - \eta k^2$ and $\mbox{Im}\,p = -\alpha k$,
that is, there are only decaying solutions of \eq{example},
which are in general oscillatory.
Let us next take the memory effect into account.
We assume now that
$\meanemf_y (t) = \alpha \meanB_x (t - \tau)$
with a positive time $\tau$
and, thinking of not too rapid changes of $\meanB_x$ during the time interval $\tau$,
express this by $\meanemf_y = \alpha (1 - \tau \partial_t) \meanB_x$.
With this relation for $\meanemf_y$ and \eq{example}, we find
$\mbox{Re}\,p = - (\eta - \alpha^2 \tau) k^2 / \big(1 + (\alpha k \tau)^2\big)$
and $\mbox{Im}\,p = - \alpha k (1+ \eta k^2 \tau) /\big(1 + (\alpha k \tau)^2\big)$.
That is, for $\alpha^2 \tau > \eta$ we have {\it growing} oscillatory solutions.
Now looking at $\hat\alpha$ in \eqref{KH33} and considering that $-\ii\omega$ can be replaced by $p$,
we see that for small $p$ it can be approximated as $\hat\alpha = \hat\alpha_0 (1 - \tau p)$,
with $k$ dependent $\hat\alpha_0$ and $\tau>0$.
Crossing over to the time domain, replacing $p$ by $\partial_t$, it becomes clear that $\hat\alpha$
indeed contains a memory effect, so the dynamo efficacy of flow II can with full right be attributed to it.

\section{Dynamo action from flows II and III}

In what follows we assume for simplicity always $v_0=w_0$ as far as numerical results
are concerned.

\subsection{Stability diagram from DNS}
\label{stabDNS}
To make progress in studying mean field dynamo action for flows II and III
beyond SOCA, we now turn to numerical solutions of \Eq{KH01}.
We discretize them
on a three-dimensional mesh in a cuboid domain employing
sixth-order finite differences in space and a third-order accurate
time-stepping scheme using the publicly available
{\sc Pencil Code}\footnote{http://pencil-code.googlecode.com/}.
In the $x$, $y$ and $z$ directions the cuboid is given
by the dimensions $2\pi/k_0$, $2\pi/k_0$, $2\pi/k$,
where $k$ defines the minimum possible wavenumber of a mean field.
The boundary conditions are always periodic in all three directions.

\begin{figure}\begin{center}
\includegraphics[width=\columnwidth]{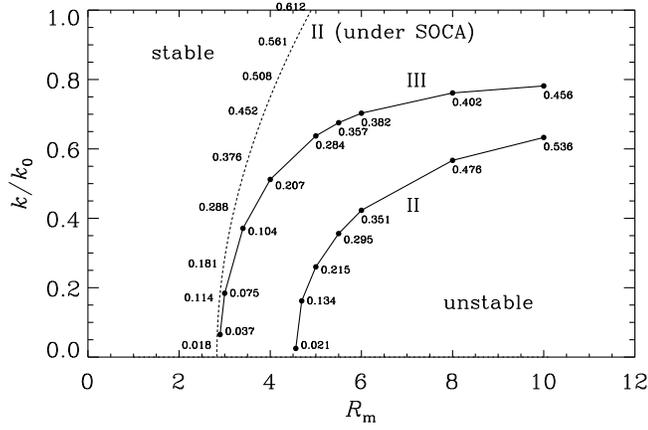}
\end{center}\caption[]{
Stability diagram for flows II and III.
The corresponding result for flow II under SOCA is shown by the dotted line.
(There is no dynamo action under SOCA for flow III.)
The numbers at the curves indicate the oscillation frequency $\omega$ in units of $v_0 k_0$.   
}\label{pdisperII_etak0}\end{figure}

\begin{figure}\begin{center}
\includegraphics[width=\columnwidth]{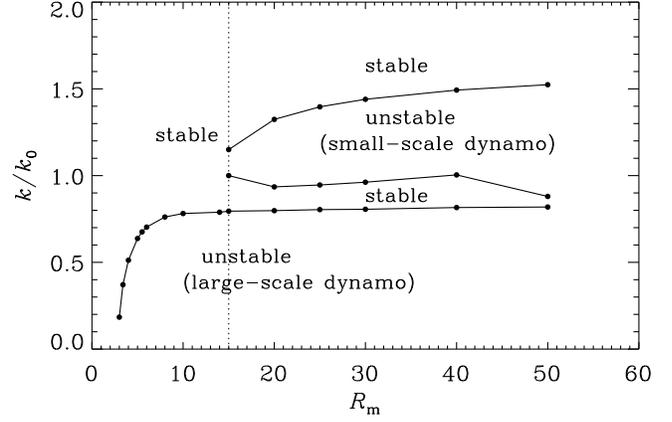}
\end{center}\caption[]{
Stability diagram for flow III showing the margins of large scale and small scale dynamo action.
The vertical dotted line at $\Rm=15$ indicates the smallest value
for which we have observed small-scale dynamo action.
}\label{DNS_kdep_eta005_III}
\end{figure}

For a given value of $k$, we determine a critical value $\Rmc$
such that there are growing solutions for $\Rm>\Rmc$, but only
decaying ones for $\Rm<\Rmc$.
In \Fig{pdisperII_etak0} we show the resulting stability
margins for flows II and III in the $k$--$\Rm$ plane;
see also \Fig{DNS_kdep_eta005_III} for higher $\Rm$ in flow III.
For comparison, we also show the corresponding result from SOCA,
where growing solutions are suggested only for flow II.
As mentioned above, the resulting stability line is already outside the
domain of validity of SOCA.

In the limit $k\to0$, we find
$\Rmc\approx4.58$ and $\approx2.9$ for flows II and III, respectively.
For $\Rm\le10$, growing solutions are only possible for $k/k_0\la0.64$
and $\la 0.78$, respectively.
Note that for $\Rm\le10$, in contrast to dynamos with flows I and IV,
growing solutions are ruled out in cubic domains, that is for $k=k_0$.
Instead, the $z$ extent of the computational domain must be larger than
the horizontal extents.
On the other hand, the limit $k\to0$ is difficult to perform numerically,
because the growth rate vanishes at $k=0$.
To study dynamos near onset, we choose $k/k_0=0.025$
so that a finite growth rate can still be easily determined.

It turns out that all solutions on the marginal lines are oscillatory.
All growing and decaying solutions encountered in determining
them are also oscillatory.
Tables~\ref{marg_dataII} and \ref{marg_dataIII} show the oscillation frequencies
for the points on the marginal curves indicated in Fig.~\ref{pdisperII_etak0}.
These tables also give the values of $\hat\alpha$, $\hat\etat$, and the resulting
growth rates $p$ obtained using the test-field method (TFM) explained in \Sec{TFM} below.

\begin{figure*}
\begin{tabular}{@{\hspace{-.06 \textwidth}}c@{\hspace{-.17 \textwidth}}c@{\hspace{-.17 \textwidth}}c}
  \hspace*{-6mm}$b_x$   & \hspace*{-4mm}$b_y$ & \hspace*{-4mm}$b_z$ \\[-1mm]
\hspace*{-.08\textwidth}\includegraphics[width=.38\textwidth]{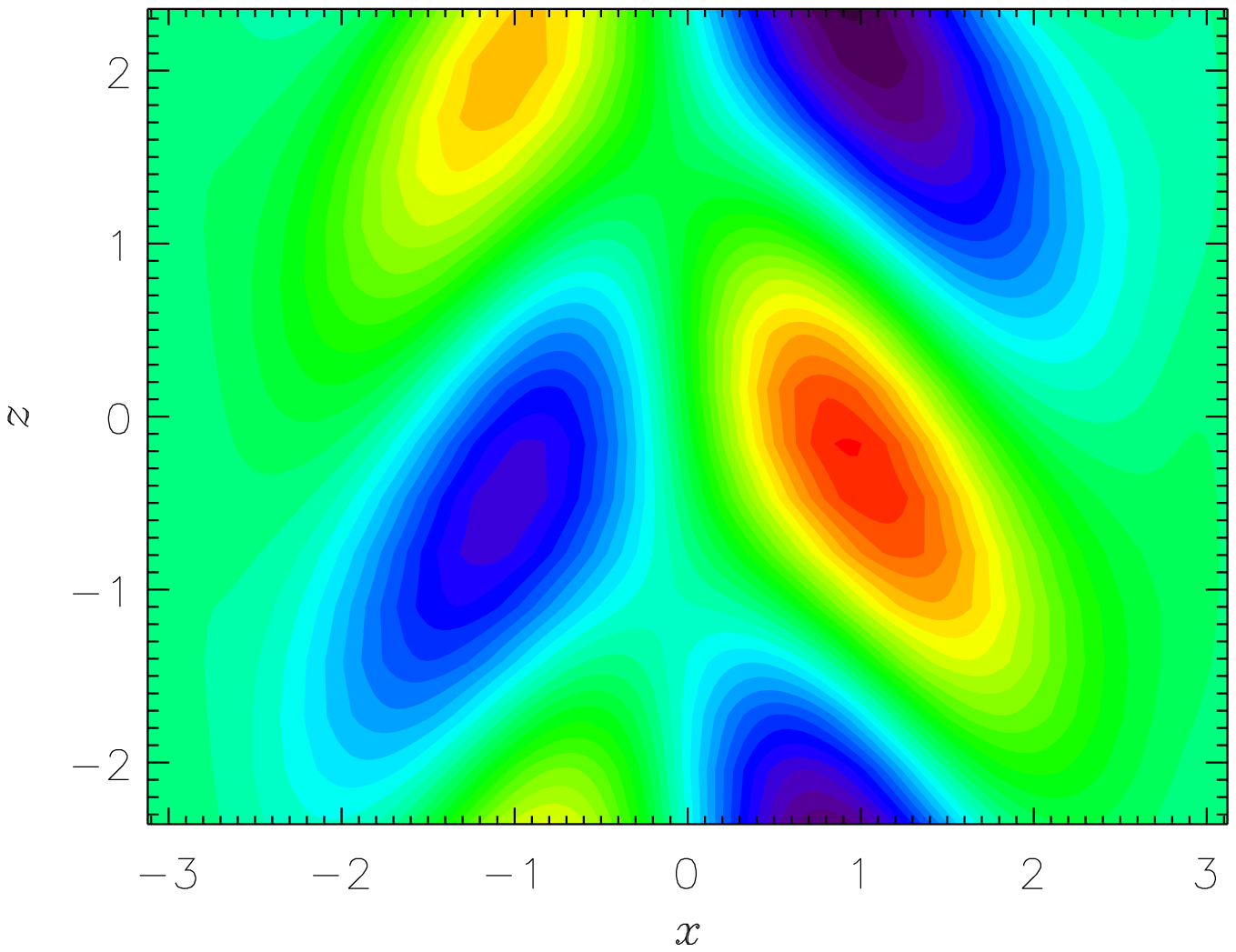}
&
\hspace*{-.07\textwidth}\includegraphics[width=.38\textwidth]{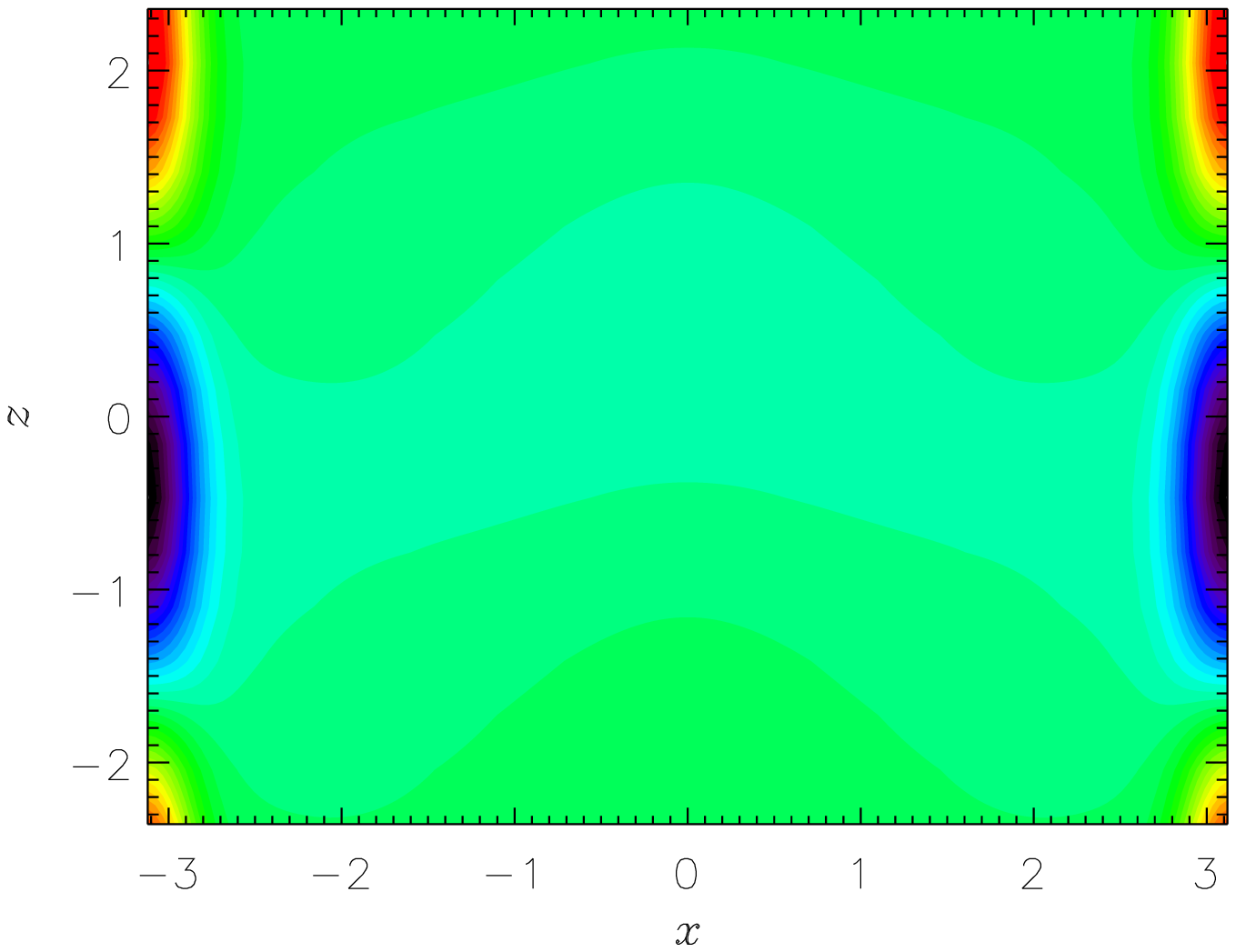}
&
\hspace*{-.07\textwidth}\includegraphics[width=.38\textwidth]{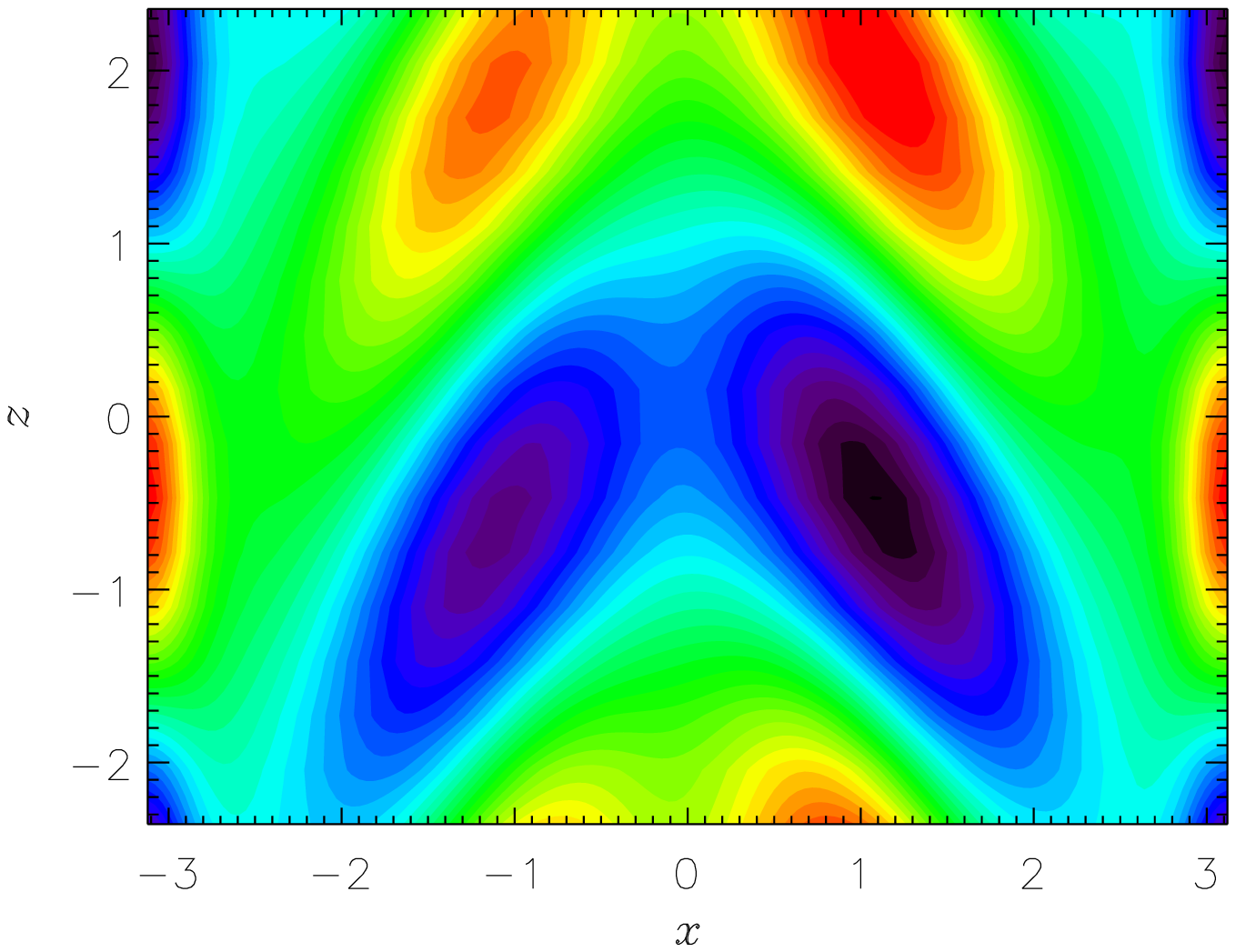}
\\[-7mm]
\includegraphics[width=.51 \textwidth]{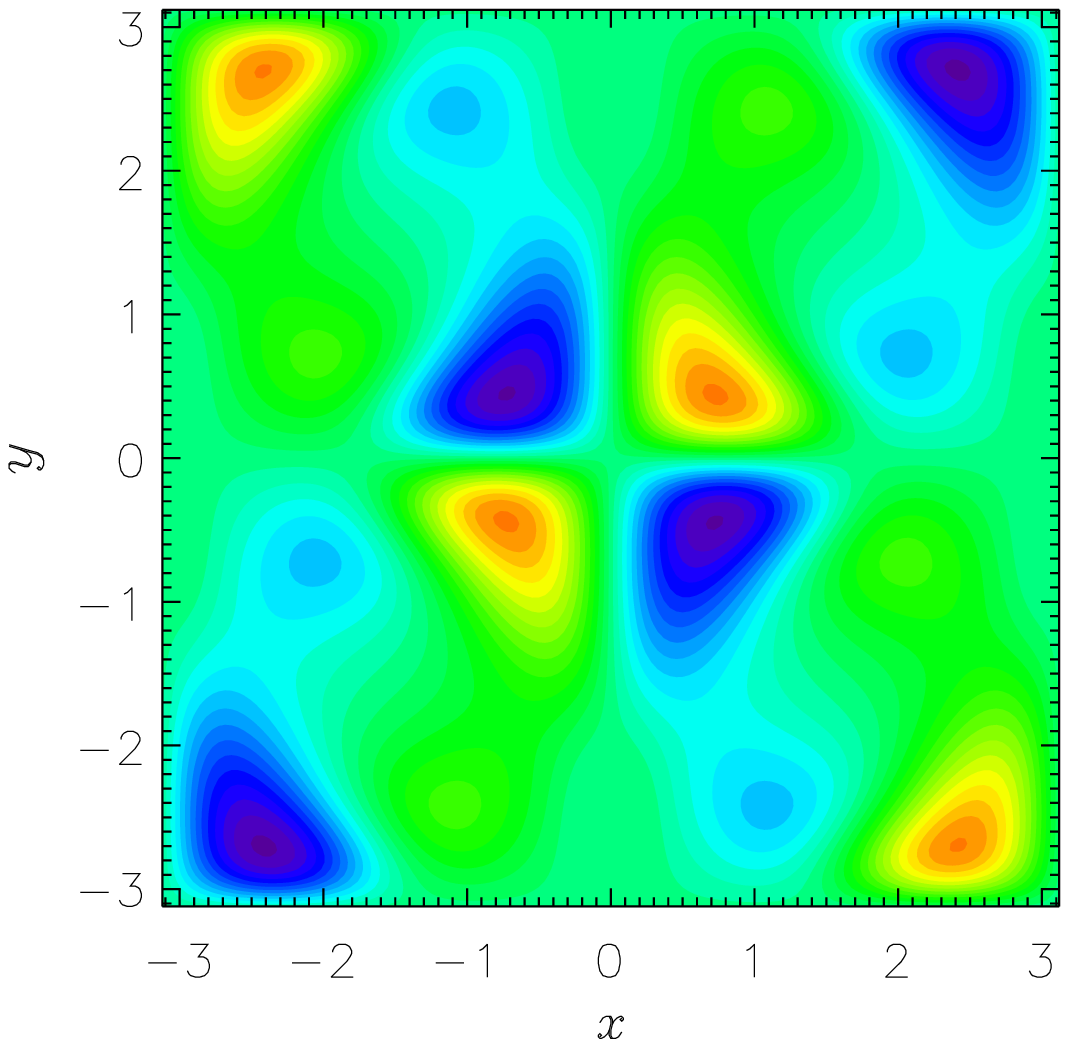}\hspace*{-.51 \textwidth}\includegraphics[width=.51 \textwidth]{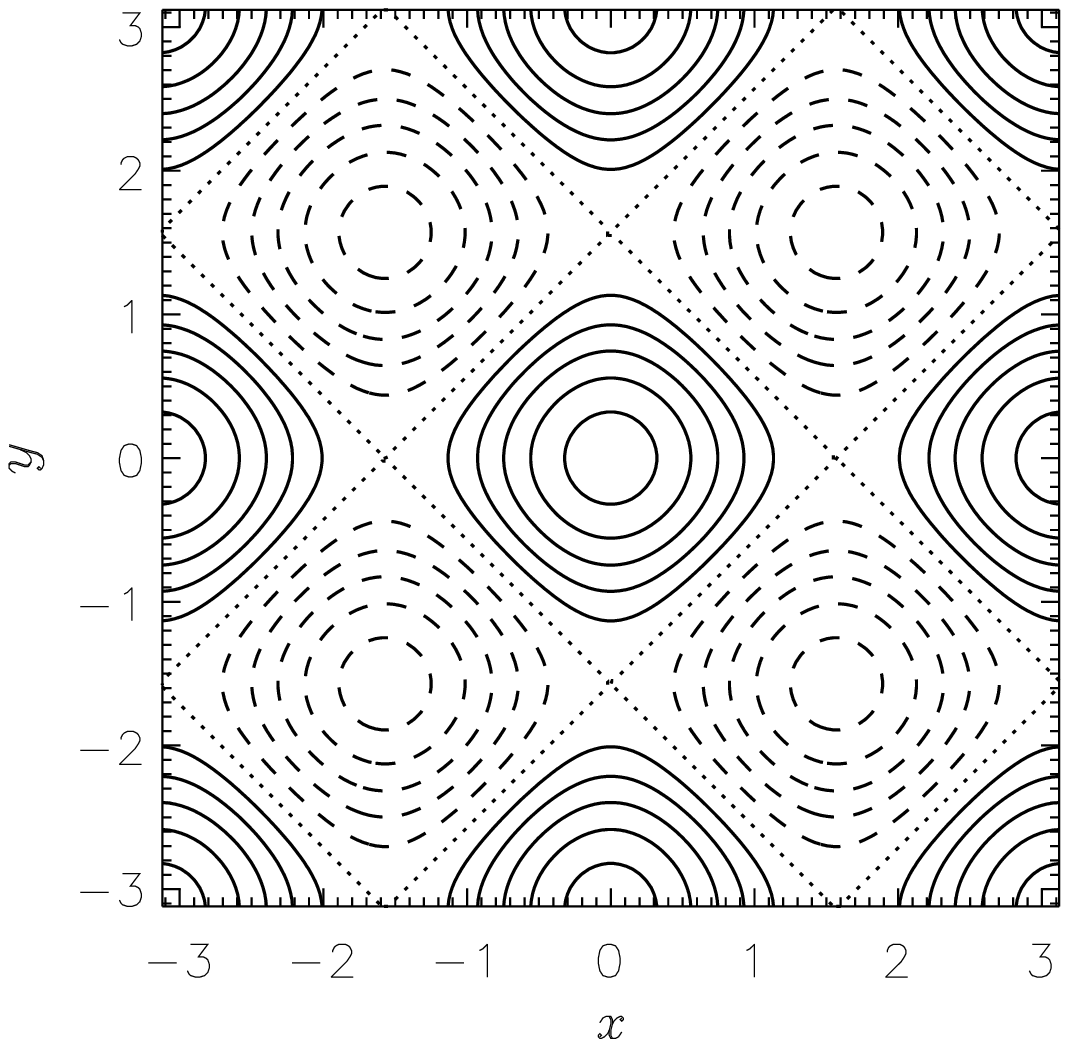} &
\includegraphics[width=.51 \textwidth]{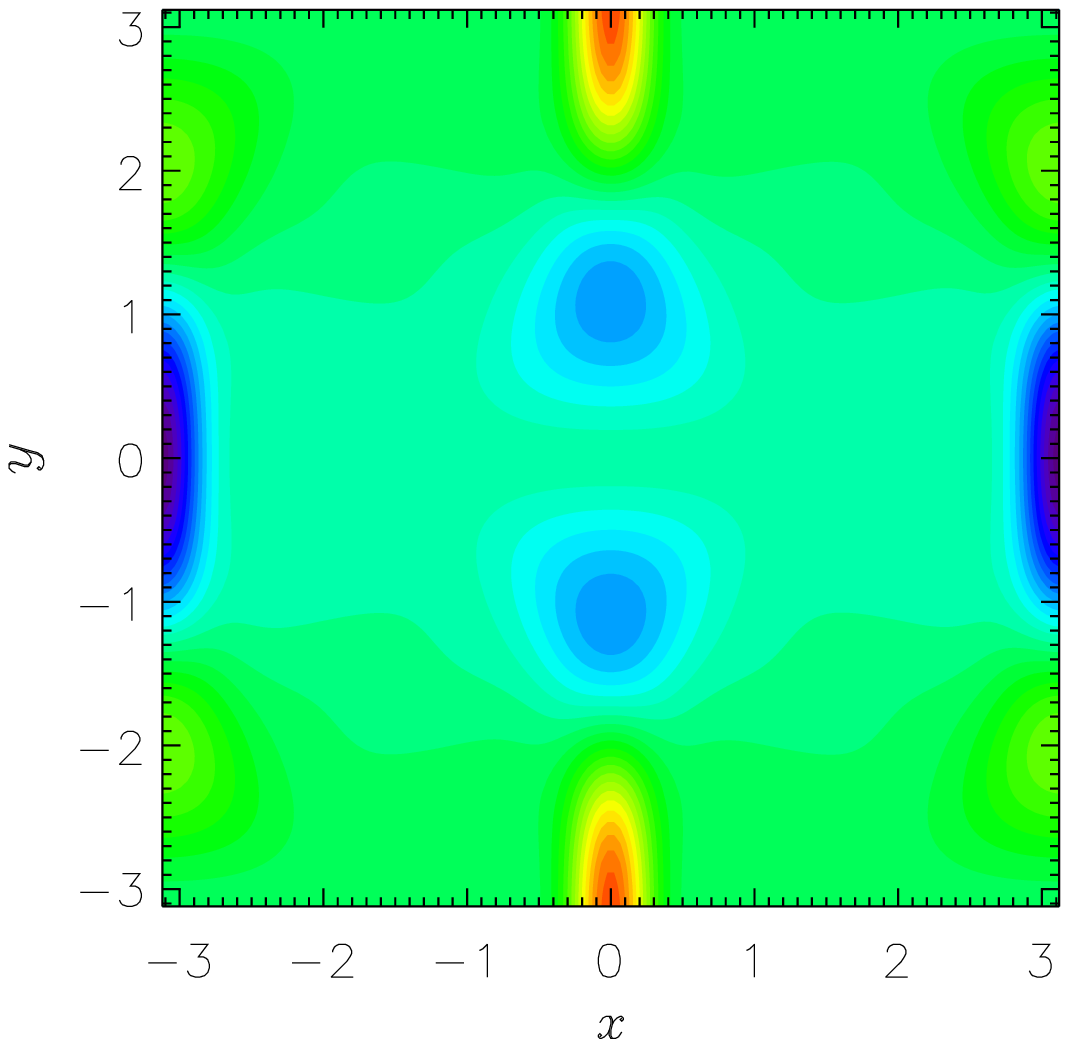}\hspace*{-.51 \textwidth}\includegraphics[width=.51 \textwidth]{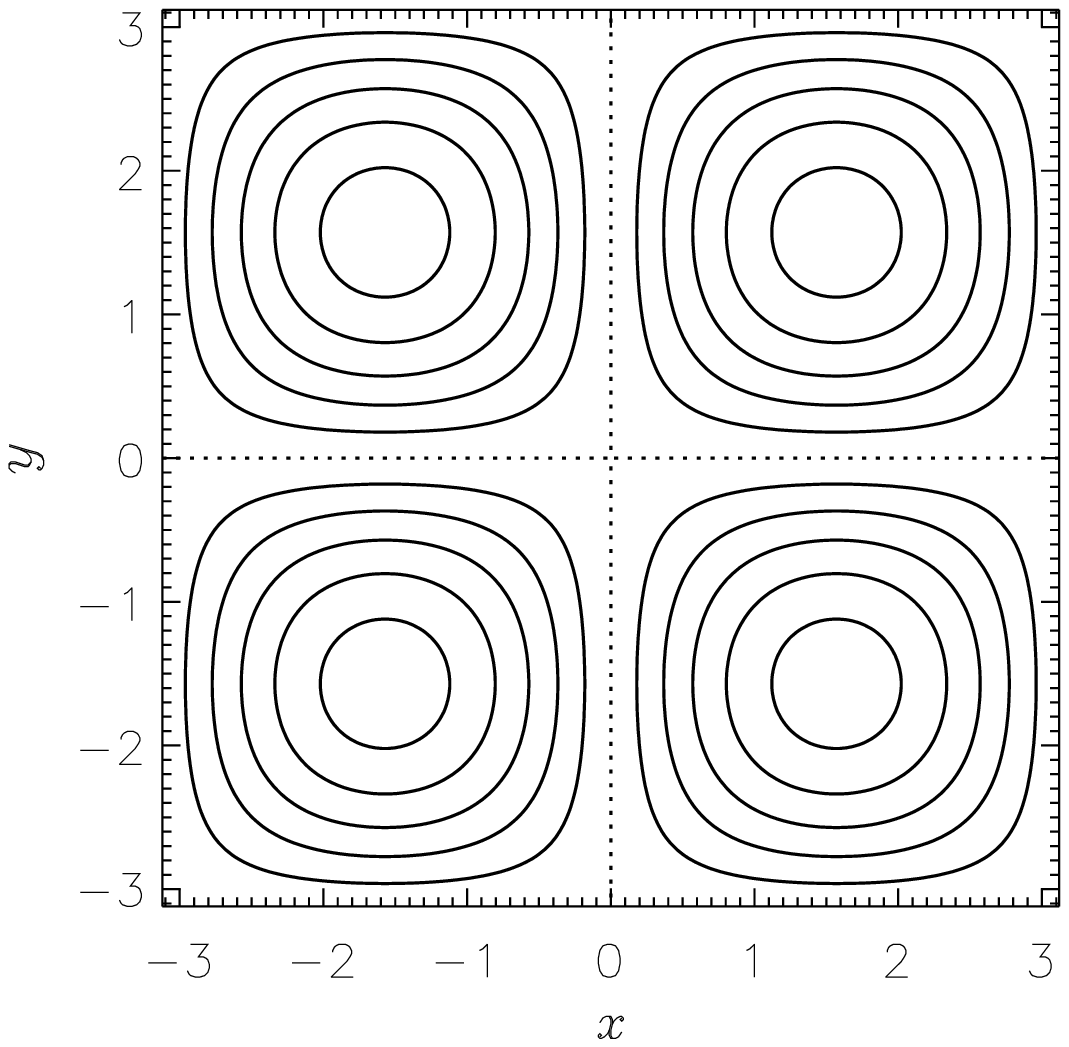}&
\includegraphics[width=.51 \textwidth]{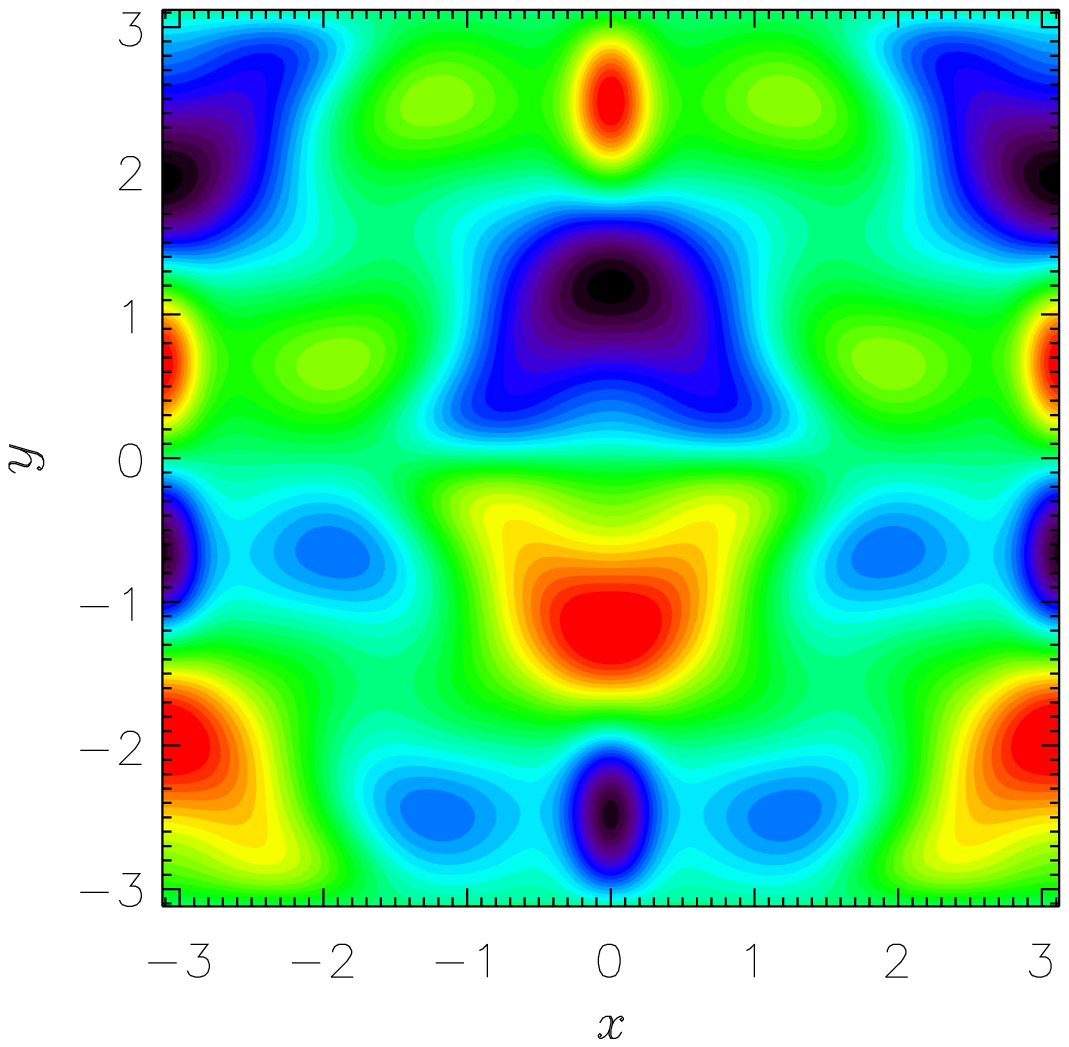}\hspace*{-.51 \textwidth}\includegraphics[width=.51 \textwidth]{RobertsIII_uuxy_xy}\\[-3mm]
\end{tabular}
\caption{\label{small-scale}
Small-scale dynamo for $\Rm=25$ and $k=1.25$.
Left, middle, right: isolines of $b_x$, $b_y$ and $b_z$, respectively.
Top, bottom: planes $y=0$ and $z=0$, respectively.
In the lower left panel, isolines of $u_z$ are overplotted
and in the lower middle and right ones streamlines of $u_{x,y}$.
}\end{figure*}

For flow III,  a second $k$ interval with dynamo action is observed for $\Rm\gtrsim15$
(see \Fig{DNS_kdep_eta005_III}), where in the DNS
initially both the mean and the total fields are decaying to very low values. While $\Brms$ starts to grow again at $t\approx 90 (v_0 k_0)^{-1}$, $\meanBB$ continues to fall.
However, at $t\approx 170 (v_0 k_0)^{-1}$, many orders of magnitude below $\Brms$,  also $\meanBB$ starts to grow again
and the growth rates of $\Brms$ and $\meanBB$ turn out to be equal. Moreover, both
the total $\BB$ and $\meanBB$ are oscillatory with the same frequency,
differing though from the one detected in $\meanBB$ during the initial decay.
As $\BB$ is clearly dominated by $\bb$, we may identify the growing field as a small-scale dynamo mode,
given that the horizontal scales of $\bb$ and $\uu$ are the same albeit the vertical scale of $\bb$ is just the same as that of $\meanBB$; see \Fig{small-scale}.
Regarding the nature of the growing $\meanBB$, see the discussion at the end of \Sec{pumping}.

\begin{table}
\caption{\label{marg_dataII} Oscillation frequency, transport coefficients and resulting complex growth rate $p$ according
to \eqref{KH53} for the points $(\Rmc,k_{\rm crit})$ on the marginal curve for flow II, see Fig.~\ref{pdisperII_etak0};
$\tilde{k}_{\rm crit}=k_{\rm crit}/k_0$, $\widetilde{\omega}=\omega/(v_0 k_0)$.
 }
\begin{tabular}{@{\hspace{0cm}}r@{\hspace{2mm}}c@{\hspace{2mm}}l@{\hspace{1.5mm}}l@{\hspace{1.7mm}}l@{\hspace{.8mm}}r@{\hspace{0cm}}}
\hline
\hline
$\!\!\!\Rmc$ & $\tilde{k}_{\rm crit}$ & $\quad\widetilde\omega$  & $\quad\;10\hat{\alpha}/v_0$  & $10\hat{\eta}/(v_0/k_0)$  &  $\; p/(v_0 k_0)\quad$ \\
\hline
  4.56  &  0.025  &    0.021 & $8.310 \!-\!0.197\ii $&$5.628 \!-\!0.094\ii$&$      4\!\times\!10^{-6}\!+\!    0.021\ii $ \\
  4.69  &  0.162  &    0.134   & $ 8.177\!-\!1.268\ii$ &$5.688 \!-\!0.627\ii$&$      3\!\times\!10^{-5}\!+\!   0.134\ii     $ \\
  5.00  &  0.260  &    0.215  &  $8.009 \!-\!2.034\ii $&$5.819 \!-\!1.032\ii$&$      3\!\times\!10^{-5}\!+\!   0.215\ii     $ \\
  5.50  &  0.356  &    0.295  &  $7.757 \!-\!2.803\ii $&$6.051 \!-\!1.491\ii$&$      6\!\times\!10^{-5}\!+\!   0.295\ii     $ \\
  6.00  &  0.423  &    0.351  &  $7.497 \!-\!3.355\ii $&$6.280 \!-\!1.904\ii$&$    -3\!\times\!10^{-4}\!+\!    0.351\ii    $  \\
  8.00  &  0.567  &    0.476  &  $6.441 \!-\!4.666\ii $&$6.977 \!-\!3.443\ii$&$      7\!\times\!10^{-5}\!+\!     0.476\ii   $ \\
10.00  &  0.633  &    0.536  &  $5.448 \!-\!5.219\ii $&$7.244 \!-\!4.784\ii$&$      4\!\times\!10^{-5}\!+\!    0.537\ii   $ \\
\hline
\end{tabular}
\end{table}

\begin{table}
\caption{\label{marg_dataIII} As Table \ref{marg_dataII}, but for flow III,  hence $p$ from \eqref{disperIII};
$\tilde{k}_{\rm crit}=k_{\rm crit}/k_0$, $\widetilde{\omega}=\omega/(v_0 k_0)$.
}
\begin{tabular}{@{\hspace{0cm}}r@{\hspace{2mm}}c@{\hspace{2.5mm}}l@{\hspace{1.5mm}}l@{\hspace{1.5mm}}l@{\hspace{1.mm}}r@{\hspace{0cm}}}
\hline
\hline
$\!\!\!\Rmc$ & $\tilde k_{\rm crit}$ & $\quad \widetilde\omega$  & $\quad10 \hat{\gamma}/v_0$  & $\:10\hat{\eta}/(v_0/k_0)$  &  $\quad p/(v_0 k_0)\quad$ \\
\hline
2.90   &   0.065   &  0.037 &   $ 5.70 \!-\! 0.309\ii$ &  $1.29 \!+\! 0.0219\ii$&$  9\!\times\!10^{-6} \!+\!   0.037\ii  $\\
2.94   &   0.132   &  0.075 &   $ 5.69\!-\!0.622\ii$ &  $1.31 \!+\! 0.0430\ii$&$   3\!\times\!10^{-6} \!+\!   0.075\ii  $\\
3.00   &   0.184   &  0.104   &   $ 5.68\!-\!0.861\ii$ &  $1.34 \!+\! 0.0580\ii$&$   4\!\times\!10^{-5} \!+\! 0.104\ii    $ \\
3.40  &   0.371    &  0.207   &   $ 5.62\!-\!1.65\ii$  &  $1.51 \!+\! 0.0954\ii$&$   -9\!\times\!10^{-7} \!+\! 0.207\ii       $ \\
4.00  &   0.512    &  0.284   &   $ 5.60\!-\!2.17\ii$  &  $1.73 \!+\! 0.111  \ii$&$    1\!\times\!10^{-4} \!+\! 0.284\ii       $\\
5.00  &   0.638    &  0.357   &   $ 5.69\!-\!2.57\ii$  &  $2.03 \!+\! 0.136  \ii$&$   -3\!\times\!10^{-5} \!+\! 0.357\ii       $\\
5.50   &   0.676   &  0.382   &   $ 5.76\!-\!2.69\ii$  &  $2.17 \!+\! 0.153  \ii$&$  -2\!\times\!10^{-4} \!+\! 0.382\ii       $\\
6.00   &   0.703   &  0.402   &   $ 5.84\!-\!2.79\ii$  &  $2.30 \!+\! 0.170  \ii$&$   1\!\times\!10^{-4} \!+\! 0.402\ii       $\\
8.00   &   0.761   &  0.456   &   $ 6.16\!-\!3.05\ii$  &  $2.76 \!+\! 0.233  \ii$&$  2\!\times\!10^{-4} \!+\! 0.455\ii       $\\
10.00  &   0.782  &  0.488   &   $ 6.44\!-\!3.25\ii$ &   $3.15 \!+\! 0.255  \ii$&$   8\!\times\!10^{-5} \!+\! 0.488\ii    $ \\
\hline
\end{tabular}
\end{table}

\subsection{Mean-field interpretation}

\subsubsection{Test-field method}
\label{TFM}

The test-field method  (TFM)
is a tool for identifying the complete set of transport coefficients that define $\meanEEEE$  for a given flow $\uu$.
It does not suffer from restrictions like SOCA
as the full \Eq{KH07} is solved numerically for $\bb$.
This is done for a number of different mean fields, called the {\it test fields},
which must be prescribed properly such that the wanted coefficients can be obtained unambiguously \citep{Schrin07}.
We choose here the four linearly independent fields
\EQ
\meanBB^{p\rm c}= B_0 \ee_p \,e^{-\ii\omega t} \cos kz ,\quad
\meanBB^{p\rm s}= B_0 \ee_p \,e^{-\ii\omega t} \sin kz,  \label{TFs}
\EN
$p=1,2$, with the unit vectors in $x$ and $y$ direction, $\ee_{1,2}$
and a real $\omega$ describing a frequency.
Since the flow is steady, we can solve for the time dependence in
Fourier space by assuming the solutions to be proportional to
$e^{-\ii\omega t}$, that is, purely oscillatory.
As mentioned above, we may also employ the Laplace transform,
then replacing $-\ii\omega$ by the complex time increment $s=\lambda-\ii\omega$.
\EEq{KH07} thus results in the following system for
the real and imaginary parts of the complex amplitude of $\bb$,
$\hat\bb(\xx,s)=\hat\bb\upre+\ii\,\hat\bb\upim$ (cf.\ \Eq{KH31})
\EQ
\begin{aligned}
\eta \nab^2 \hat\bb\upre + \nab \!\times\! (\uu \!\times\! \hat\bb\upre)'
- \lambda \hat\bb\upre - \omega \hat\bb\upim
&= - \nab \!\times\! \big(\uu \!\times\! \hat{\meanBB}\upresh{4}\big) \, ,  \\
\eta \nab^2 \hat\bb\upim+ \nab \!\times\! (\uu \!\times\! \hat\bb\upim)'
- \lambda \hat\bb\upim + \omega \hat\bb\upre
&= - \nab \!\times\! \big(\uu \!\times\! \hat{\meanBB}\upimsh{4}\big) \, ,
\end{aligned}
\label{KH07ri}
\EN
where $\meanBB$ is any out of the set defined by \eqref{TFs}.
In general we then determine coefficients $\hat\alpha_{ij}(z,k,s)$ and
$\hat\eta_{ij}(z,k,s)$ such that they obey the equations
\EQ
\hat\meanemf_i^{\,pq}\!\!=
\hat\alpha_{ij}\hat\meanB_j^{\,pq}\!\!
-\hat\eta_{ij}\mu\hat\meanJ_j^{\,pq}\!\!, \label{tfsys}
\EN
where $i,j,p=1,2$, the superscript $q$ is either c or s,
$\mu\hat\meanJ_j^{\,pq}=\nab\times\hat\meanB_j^{\,pq}$,
and a common argument $(z,k,s)$ on all functions has been dropped.
Given that we are asking for eight coefficients, $\hat\alpha_{ij}$ and $\hat\eta_{ij}$, and that the test fields
are linearly independent, the system \eqref{tfsys} is just sufficient to yield a unique result.
For the $z$ invariant Roberts flows the coefficients are also independent of  $z$, hence we will drop this
argument in the following.

\subsubsection{Test-field results for flow II}

For a first verification of the TFM we have calculated the transport coefficients
for the points $(\Rmc,k_{\rm crit})$ on the marginal curve of Fig.~\ref{pdisperII_etak0}
employing $k_{\rm crit}$ and the detected oscillation frequency in \eqref{KH07ri}.
It turned out that to high accuracy, $\hat{\alpha}_{ij}$ and $\hat{\eta}_{ij}$
have the same structure as obtained under SOCA, that is,
$\hat{\alpha}_{12}=\hat{\alpha}_{21}=\hat{\alpha}$,
$\hat{\eta}_{11}=\hat{\eta}_{22}=\hat{\eta}$ with all other components vanishing.
When inserting the results in the thus valid dispersion relation \eqref{KH53} the outcome should be $p=0 - \ii \omega$.
Indeed this was confirmed with high accuracy, see Table~\ref{marg_dataII}, where we list
$\hat{\alpha}$ and $\hat{\eta}$ along with  $p$ obtained from them.
Note that the marginal points were determined by an iterative procedure and their oscillation frequency by a fit, so the achievable agreement of the two results for $p$ is limited already by the quality of the input data to the TFM.
\begin{figure}\begin{center}
\includegraphics[width=\columnwidth]{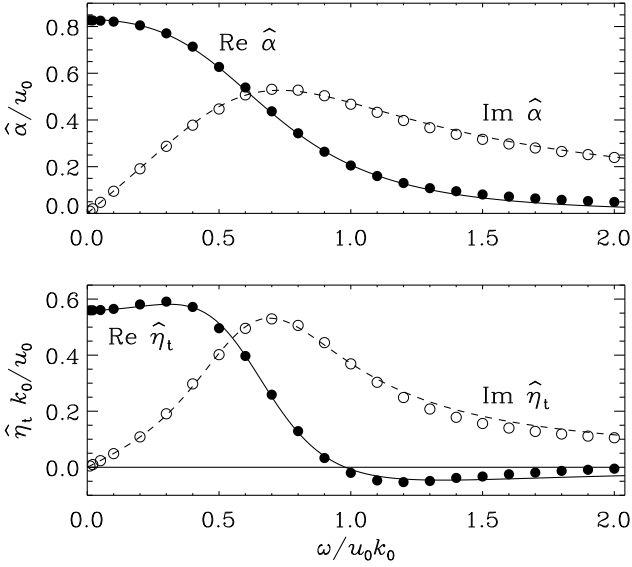}
\end{center}\caption[]{
$\omega$ dependence of
$\hat\alpha(k, \omega)$ and $\etath(k, \omega)$
for flow II with $k/k_0=0.025$ and $\Rm=4.6$.
}\label{presults}\end{figure}

\begin{figure}\begin{center}
\includegraphics[width=\columnwidth]{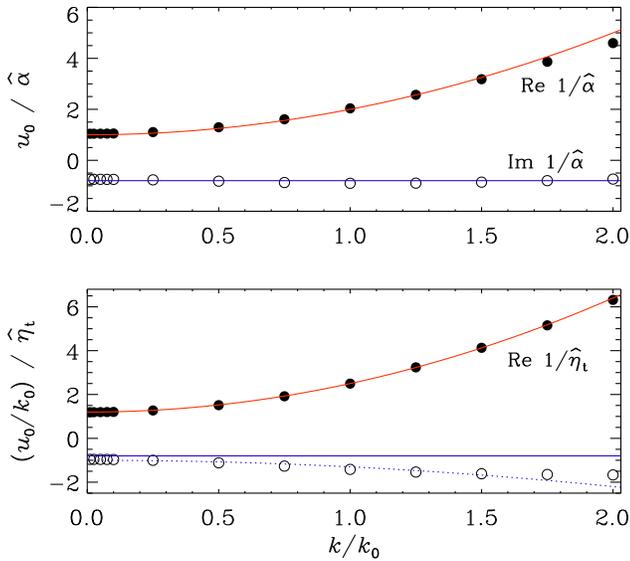}
\end{center}\caption[]{
Dependence of $1/\hat\alpha(k,\omega)$ and $1/\etath(k,\omega)$ on $k$
for flow II with $\omega/v_0k_0=0.5$ and $\Rm=4.6$.
The Lorentzian fits are obtained with $k_0\ell=1.0$ for
$\rm{Re}\, \hat\alpha$ and with $1.14$ for $\rm{Re} \,\etath$, respectively.
In the second panel, $\mbox{Im}\,\etath$ is not a constant
(blue solid line) and a Lorentzian with $k_0\ell=0.55$ fits better (blue dotted line).
}\label{presults_k_T32_Rm4p6_om05_kscan_nosoca}\end{figure}

Next, we use the TFM to study the $\omega$ dependence
of the transport coefficients in the neighborhood
of the lowest point on the marginal curve,
$k/k_0=0.025$ and $\Rm\approx4.56$, but fixing $\lambda=0$ in \eqref{KH07ri}.
For simplicity we write $\hat\alpha(k,\omega)$ and $\etath(k,\omega)$,
dropping the imaginary unit in the second argument.
The results are shown in  \Fig{presults}

These data can be utilized to infer the dependences of the coefficients on
the {\it complex} increment $s$
which opens the way for predicting growth (or decay)
rates also for points in the $\Rm$--$k$ plane distant from the marginal curve.
To accomplish this, we have to find an approximation of $\hat{\alpha}$ and $\hat{\eta}$
as analytic functions of $\ii \omega$, in which we are allowed to replace $\ii\omega$ subsequently
by $s$.
Employing these functions in \Eq{KH53} with $s=p$, enables us to solve consistently for $p$.

For small values of $\omega$, the resulting functions $\hat{\alpha}$ and $\hat{\eta}_{\rm t}$
are proportional to $(1-\ii\omega\tau)^{-1}$, in qualitative agreement with the SOCA result \Eq{KH33}.
However, the values of $\tau$ are no longer the same for $\hat\alpha$ and $\etath$.
For larger values of $\omega$, the resulting $\omega$ dependences
become more complicated and can be fitted to expressions of the form
\EQ
\begin{aligned}
\hat\alpha(s) &=\alpha_0\frac{1+\tau_{\alpha 1} s}{1+\tau_\alpha s-(\tau_{\alpha 2}s)^2},\\
\etath(s)     &=\eta_{t0}\frac{1+\tau_{\eta 1} s}{1+\tau_\eta s-(\tau_{\eta 2}s)^2},
\end{aligned}
\label{fit}
\EN
where $s=-\ii\omega$ with real coefficients $\tau_\ast$, $\alpha_0$,
$\eta_{t0}$; see Table~\ref{Ttimescale}.
Note that $\hat\alpha$ and $\etath$ are real only when $s$ is so.  
The result is shown in \Fig{presults} as continuous lines.

\begin{table}\caption{
Parameters of the fits \eqref{fit} to the data points shown in \Fig{presults}.
Here, $\alpha_0$ is normalized by $v_0$, $\eta_{t0}$ by $v_0/k_0$,
and all $\tau_\ast$ by $(v_0 k_0)^{-1}$.
Normalization is indicated by tildes.
}\vspace{12pt}\centerline{\begin{tabular}{l|cccc}
\hline
\hline
$\sigma$ &  $\widetilde\sigma_0$ & $\widetilde\tau_{\sigma 1}$ & $\widetilde\tau_\sigma$ & $\widetilde\tau_{\sigma 2}$ \\
\hline
$\alpha$ & 0.83 & 0.904 & 2.05 & 1.296 \\
$\eta$   & 0.56 & 0.643 & 1.49 & 1.414 \\
\hline
\label{Ttimescale}\end{tabular}}\end{table}

We have also identified the $k$ dependence of $\hat\alpha$ and $\etath$ which, for $\omega=0$,
prove to be roughly compatible with that of a Lorentzian, $\big(1+(k\ell)^2 \big)^{-1}$,
again in qualitative agreement with the SOCA result \Eq{KH33},
but with different values $\ell_\alpha$ and $\ell_\eta$ of $\ell$ for $\hat\alpha$ and $\etath$.
In \Fig{presults_k_T32_Rm4p6_om05_kscan_nosoca} we show the result
for $\Rm=4.6$ and the arbitrarily chosen value $\omega/v_0k_0=0.5$,
where the fits (overplotted lines) are obtained with
$k_0\ell_\alpha=1.0$ and $k_0\ell_\eta=1.14$.
Note that the SOCA result \Eq{KH33} for $\omega\ne0$ suggests $k$ independent imaginary parts
of $1/\hat{\alpha}$ and $1/\etath$.
From \Fig{presults_k_T32_Rm4p6_om05_kscan_nosoca} one can see
that this is well satisfied for $1/\hat{\alpha}$, but not for $1/\etath$.

\subsubsection{Self-consistent growth rate from test-field method}

In order to predict the growth rate at a given point in the $\Rm-k$ plane,
one could proceed as exemplified above: determine the $\omega$ dependence
of $\hat\alpha$ and $\etath$, establish analytical approximations for
$\hat\alpha(s)$ and $\etath(s)$, $s=-\ii\omega$, via a fit procedure,
employ them in \Eq{KH53} with $s=p$ and finally solve for $p$.

A less cumbersome way is offered by an iterative approach
defined schematically by
\EQ
\hspace{-1cm}\begin{aligned}
&p_0 = \,\text{initial guess} \\
&\text{do while stop criterion}\ne\text{TRUE} \\
&\quad\hat\alpha(p_n),\etath(p_n) := \text{TFM}(p_n) \\
& \quad p_{n+1} := p\big(\hat\alpha(p_n),\etath(p_n)\big)   \\
& \quad n:=n+1\\
& \text{enddo}
\end{aligned}\label{iter}
\EN
where TFM$(p_n)$ stands for the application of the TFM, see \Sec{TFM}, with the complex $p_n$ as input
and $p(\hat\alpha,\etath)$ for the rhs of the dispersion relation \Eq{KH53}.
Of course both major steps in \eq{iter} have to be carried out with the
chosen $\Rm$ and $k$ and an appropriate stop criterion has to be applied.

\begin{figure}\begin{center}
\includegraphics[width=\columnwidth]{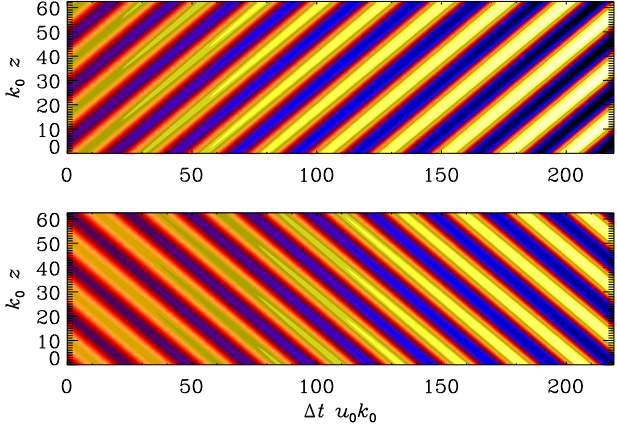}
\end{center}\caption[]{
$\meanB_x$ and $\meanB_y$ in a $zt$ diagram as obtained from
DNS for $\Rm=6$, $k/k_0=0.3$, using random initial conditions.
The growth rate is $0.00408\,v_0k_0$
and the frequency $\omega=0.2567 \,v_0 k_0$,
cf.\ the predicted values in \Tab{Tdisper}.
$\Delta t=t-t_0$ with $t_0$ defined by dominance of
the fastest growing mode for $t>t_0$.
}\label{pbutter_64x192_251_R6_ran}\end{figure}

We demonstrate this now for flow~II in the special case of $\Rm=6$ and $k/k_0=0.3$,
which is well outside the domain of validity of SOCA.
We adopt $p_0=0$ as the initial guess and obtain after seven iterations,
a four-digit converged result with growth rate
$\lambda=0.00408 v_0 k_0$ and the frequency $\omega=0.2582 v_0 k_0$.
\Tab{Tdisper} lists all iterations needed.
For comparison we have performed a DNS, again with $\Rm=6$ and random initial conditions,
and an aspect ratio of the cuboid corresponding to $k/k_0$,
which for periodic boundary conditions
allows harmonic mean fields with the desired $k$ to evolve.
We are also interested in the eigenfunction corresponding
to the fastest growing mode.
This means that the DNS has to run long enough (until $t=t_0$, say)
so that all other modes have become subdominant.
We observe indeed both components of $\meanBB$ to be growing with just the predicted growth rate and frequency,
see \Fig{pbutter_64x192_251_R6_ran}.
A corresponding experiment with non-vanishing initial conditions in only one of the components
confirms their independent growth.  

\begin{table}\caption{
Iteration steps of the procedure \eqref{iter} with
$\Rm=6$ and $k/k_0=0.3$ for flow II.
}\vspace{12pt}\begin{tabular}{@{\hspace{0.1cm}}ccc@{\hspace{0.4cm}}r@{\hspace{0.1cm}}}
$n$\!\! & $\hat\alpha/v_0$ &  $\etath/(v_0/k_0)$ & $p /(v_0 k_0)\quad\quad$ \\
\hline
1\!\! & $0.8696+0.0001\ii$ & $0.5367-0.00003\ii$ &$\!\!-0.063280-0.2609\ii$ \\
2\!\! & $0.9095+0.2747\ii$ & $0.6846-0.04088\ii$ & $0.005796-0.2765\ii$ \\
3\!\! & $0.8261+0.2698\ii$ & $0.6337-0.10560\ii$ & $0.008907-0.5730\ii$ \\
4\!\! & $0.8294+0.2487\ii$ & $0.6212-0.09044\ii$ & $0.003702-0.2570\ii$ \\
5\!\! & $0.8349+0.2501\ii$ & $0.6253-0.08747\ii$ & $0.003753-0.2583\ii$ \\
6\!\! & $0.8345+0.2514\ii$ & $0.6260-0.08857\ii$ & $0.004080-0.2583\ii$ \\
7\!\! & $0.8341+0.2513\ii$ & $0.6257-0.08874\ii$ & $0.004077-0.2582\ii$ \\
8\!\! & $0.8341+0.2513\ii$ & $0.6257-0.08874\ii$ & $0.004077-0.2582\ii$ \\
\hline
\label{Tdisper}\end{tabular}\end{table}

\begin{figure}\begin{center}
\includegraphics[width=\columnwidth]{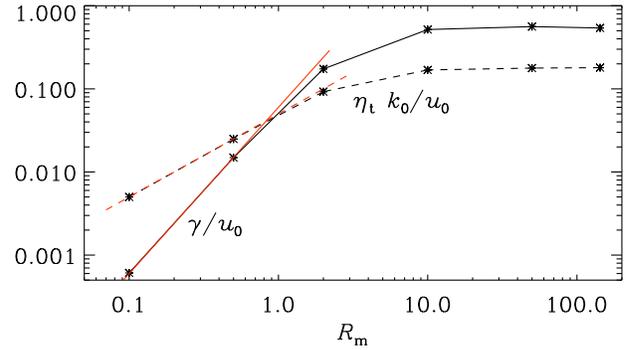}
\end{center}\caption[]{
$\Rm$ dependence of $\hat\gamma$ (solid) and $\etath$ (dashed) for flow III with $k=k_0$ and $\omega=0$.
Red lines: Scalings $\sim\Rm^2$ and $\Rm$ for $\hat\gamma$ and $\etath$, respectively.
}\label{pIII}\end{figure}

\subsubsection{Pumping effect in flow III}
\label{pumping}

We recall that, under SOCA, the dispersion relation \eq{KH53} for flow
III applies with $\hat\alpha=0$, so only decaying solutions are predicted.
However,  this is no longer true beyond SOCA:
Using the TFM, we find that for all $\Rm$, $k$ and $\omega$ considered
\EQ
-\hat\alpha_{12}=\hat\alpha_{21}\equiv\hat\gamma \ne 0,  \quad \text{still with}\quad  \hat\alpha_{11}=\hat\alpha_{22}=0,
\EN
where we have chosen the symbol $\hat\gamma$, recognizing that this effect corresponds to an advection
of the mean magnetic field with the velocity $\hat\gamma \ee$
(but without mean material transport).
This is often referred to as {\it turbulent pumping} or {\it turbulent diamagnetism}.
As for flow II, the equations for $\meanB_x$ and $\meanB_y$ decouple,
and we have here the only slightly different dispersion relation
\EQ
p = -\ii k\hat{\gamma} - (\eta + \hat{\eta}_{\rm t}) k^2
\label{disperIII}
\EN
for both $\hat{\meanB}_x$ and $\hat{\meanB}_y$.
Clearly, the pumping effect,
if acting instantaneously, that is, with a real $\hat\gamma$,
does not lead to dynamo action on its own, but gives merely rise to oscillations.
However, for $\omega\ne0$ we find always complex values of $\hat{\gamma}$
indicating the presence of the memory effect.
Like the imaginary part of $\hat\alpha$
for flow II, the one of $\hat\gamma$ has the potential to overcome
the negative real contribution to $p$ from the second term in \eqref{disperIII}.
\Tab{marg_dataIII} presents $\hat{\gamma}$ and $\etath$ for the points
on the marginal curve of flow III shown in \Fig{pdisperII_etak0}.
In the last column one finds the value of the complex growth rate obtained
when inserting the transport coefficients into \eqref {disperIII}.
As in the case of flow II, the agreement with $p$, observed in the DNS, is excellent.

\begin{figure}\begin{center}
\includegraphics[width=\columnwidth]{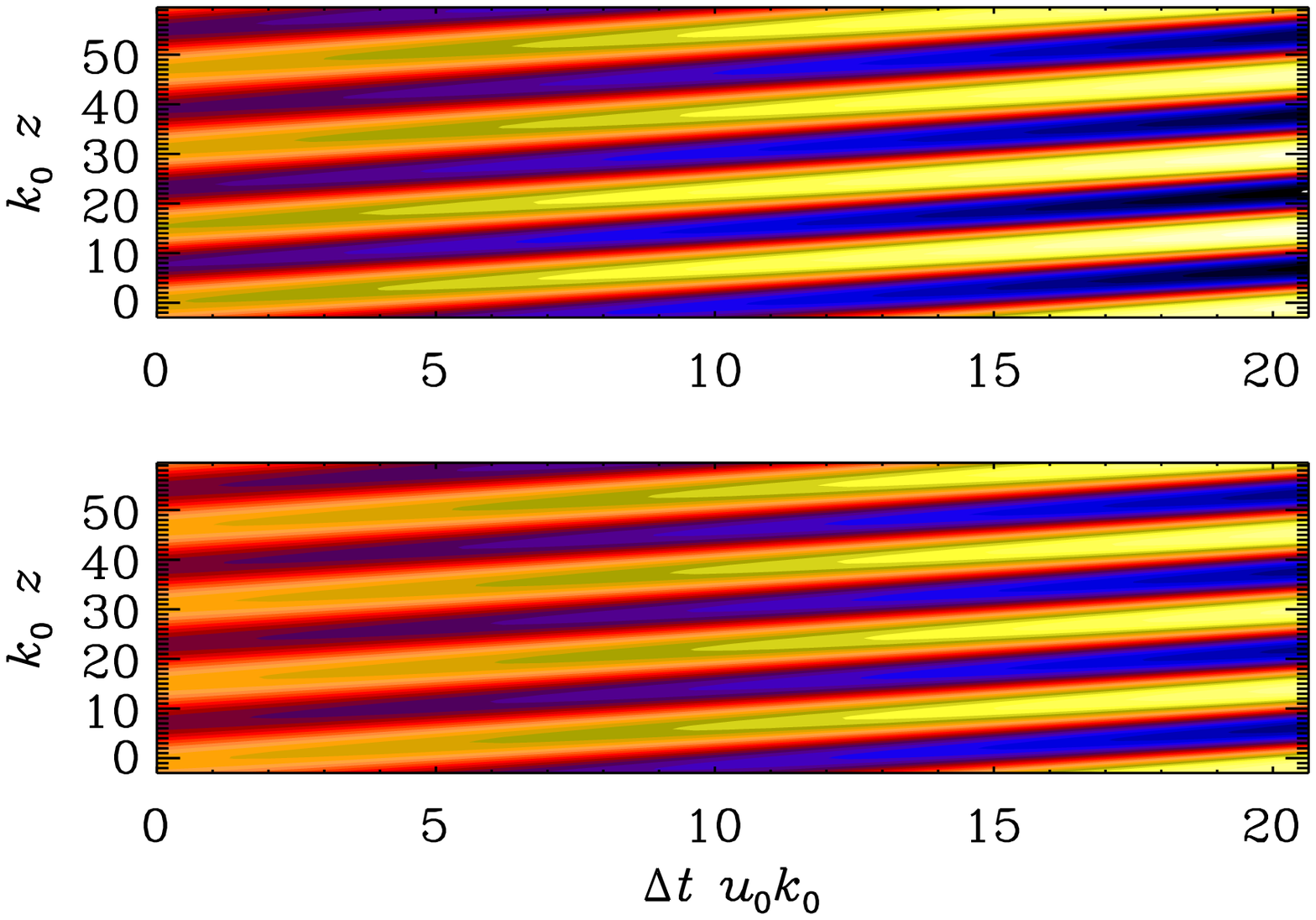}
\end{center}\caption[]{
Similar to \Fig{pbutter_64x192_251_R6_ran}, but for flow III at $\Rm=6$,
$k/k_0=0.4$.
The growth rate is $0.0471\,v_0k_0$
and the frequency $\omega=0.2877 \,v_0 k_0$,
cf.\ the predicted values in \Tab{TdisperIII}.
}\label{pbutter_III_64x192_251_R6_ran}\end{figure}

In \Fig{pIII} we show that for flow III at small values of $\Rm$, $\hat{\gamma}/v_0$ is proportional to $\Rm^2$,
which is steeper than the in general linear scaling of the components of $\alpha_{ij}/v_0$ in SOCA.
Hence, the found $\hat{\gamma}$ cannot be captured by this approximation.
We recall here a related result for the Galloway--Proctor flow \citep{RB09}, where
$\gamma/\urms$
turned out to be proportional to $\Rm^5$.
Just like for flow II, we can determine self-consistent values of $\hat{\gamma}$ and $\etath$
for given $k$ and $\Rm$ in an iterative manner such that they obey
the dispersion relation \eq{disperIII}.
As demonstrated in \Tab{TdisperIII} for $\Rm=6$ and $k/k_0=0.4$
the procedure converges,
but requires somewhat more steps than for flow II.
The normalized growth rate and frequency
resulting from the dispersion relation are
$0.04711$ and $0.2892$, respectively,
and are in very good agreement with the result of DNS;
see \Fig{pbutter_III_64x192_251_R6_ran}.

Naturally, the question arises how the polar vector $\gamma \ee$ can be constructed from any directions detectable in $\uu$.
Superficially, there seems to be only one such direction, namely just that of $\ee$, but no preferred sense of it (up or down)
is identifiable.
Indeed, from this argument one can correctly conclude that flow I does not show a pumping effect.
This is possible, because for this flow
all second-rank transport tensors can be shown to be symmetric about the $z$ axis under the planar average adopted here \citep{Rae02}.
In contrast, flow III does not show
the underlying symmetry
property.
Consequently, it
can imprint preferred directions different from $\ee$ into its relevant averages
and has therefore the potential of showing a pumping effect.
Indeed, with the vorticity $\oo=\nab\times\uu$,
a polar vector can be constructed as $\overline{\oo\times(\oo\times\uu)}$,
having only a non-vanishing $z$ component equal to $-k_0^2 v_0^2 w_0/2$.
This finding also supports the quadratic scaling of $\gamma/v_0$ with $\Rm$
for $v_0=w_0$.

It remains to clarify the nature of the mean field that grows along with the small-scale dynamo mode described in \Sec{stabDNS}.
For that, we have applied the TFM with the relevant values of $\Rm$ and $k$ as well as the growth rate and frequency measured in the DNS.
Subsequently employing the obtained transport coefficients
in the dispersion relation \eqref{disperIII}
yields a prediction of decay instead of growth,
along with a frequency differing from the one used as input to the TFM.
An attempt to determine the complex growth (or decay) rate of $\meanBB$ consistently
by the iterative method fails due to lack of convergence.
We conclude that the growing $\meanBB$ is not an eigenmode,
but enslaved by the growing $\bb$.
The only possible cause seems to be a non-vanishing
$\meanEEEE_0\equiv\overline{\uu\times\bb_0}$
where $\bb_0$ stands for the small-scale field
which would evolve in the absence of the mean field.
$\meanEEEE_0$ represents an inhomogeneity in the equation governing
$\meanBB$ and, as $\uu$ is stationary, both $\bb$ and $\meanBB$
would have the same temporal dependence as $\bb_0$
(after all transients having decayed).
We have calculated at first $\overline{\uu\times\bb}$
finding that it is more than six orders of magnitude smaller than $\urms \brms$.
Given that $\bb$ contains, along with $\bb_0$,
necessarily also a contribution from the tangling of $\meanBB$ by $\uu$,
one has to remove that part, which can be derived from the TFM values of $\hat\alpha$ and $\etath$.
The resulting $\meanEEEE_0$, although
being only a fraction of $\overline{\uu\times\bb}$ and anyway
tiny compared to $\urms \brms$, does not vanish.
However, given its decrease with increasing resolution,
we conclude that it is likely a numerical artifact.

\begin{table}\caption{
Iteration steps of the procedure \eqref{iter} with
$\Rm=6$ and $k/k_0=0.4$ for flow III.
}\vspace{12pt}\begin{tabular}{@{\hspace{0.2cm}}cccc@{\hspace{0.2cm}}}
$n$\!\! & $\hat\alpha/v_0$ &  $\etath/(v_0/k_0)$ & $p/(v_0 k_0)$ \\
\hline
1\!\! & $0.8035-0.0001\ii$ & $0.1609-0.00000\ii$&$0.05245-0.3214\ii$ \\
2\!\! & $0.8374+0.3268\ii$ & $0.2188+0.06709\ii$&$0.06905-0.3242\ii$ \\
3\!\! & $0.7018+0.2972\ii$ & $0.2189+0.01903\ii$&$0.05719-0.2777\ii$ \\
4\!\! & $0.7265+0.2543\ii$ & $0.2062+0.02417\ii$&$0.04206-0.2867\ii$ \\
5\!\! & $0.7395+0.2669\ii$ & $0.2088+0.02846\ii$&$0.04671-0.2912\ii$ \\
6\!\! & $0.7337+0.2704\ii$ & $0.2101+0.02706\ii$&$0.04788-0.2891\ii$ \\
7\!\! & $0.7330+0.2680\ii$ & $0.2095+0.02673\ii$&$0.04701-0.2889\ii$ \\
8\!\! & $0.7339+0.2680\ii$ & $0.2094+0.02699\ii$&$0.04702-0.2893\ii$ \\
9\!\! & $0.7338+0.2683\ii$ & $0.2095+0.02699\ii$&$0.04714-0.2892\ii$ \\
10\!\!& $0.7337+0.2683\ii$ & $0.2095+0.02695\ii$&$0.04711-0.2892\ii$ \\
\hline
\label{TdisperIII}\end{tabular}\end{table}

\section{Evolution equation for the mean electromotive force}

\subsection{Real space-time formulation}

So far we have demonstrated how flows II and III can be represented in
a mean-field model
if the temporal behavior of the mean field is an oscillation with exponentially growing or decaying amplitude, that is,
$\meanBB\sim \exp(pt)$ with a complex $p$.
It is highly desirable to overcome this limitation and to
allow general time behavior, e.g., when transient processes are to be considered.
This can be accomplished
by establishing analytical approximations for the $k$ and $\omega$ dependences of $\hat\alpha$ and $\etath$ and
Fourier-backtransforming them into the spacetime domain for obtaining the convolution kernels in \Eq{KH11}.
This integral representation of $\meanEEEE$ thus becomes practically handleable.

However, performing a convolution in time is cumbersome
from a numerical point of view, because one would need
to store the magnetic field at sufficiently many previous times.
Moreover, the spatial integral represents a global operation
requiring global communication in parallelized codes.
Thus a differential equation governing $\meanEEEE$ instead of
an integral one would be a major benefit.
Such a model would also open the gateway to include nonlinearities
due to magnetic quenching of the transport coefficients,
which otherwise must be kept out.

For isotropically forced turbulence, \cite{RB12} found that the kernels
of the $\alpha$ and $\etat$ tensors, which are then isotropic,
can both well be approximated in Fourier space by
\EQ
\hat\sigma = \frac{\sigma_0}{1+(k\ell)^2 - \ii\omega\tau}
\label{ansatz}
\EN
with $\sigma$ standing for $\alpha$ or $\etat$;
see also \cite{BKM04} for passive scalars.
Multiplying now
$\hat\meanEEEE = \hat\alpha \hat\meanBB -\etath \mu \hat\meanJJ$
with the
denominator of \eqref{ansatz} and returning from the $k\hspace{.2mm}\omega$
domain to the spacetime domain,
we arrive at a diffusion-type operator
acting on $\meanEEEE$ and thus at
the simple evolution equation
\EQ
\left(1-\ell^2\partial_z^2+\tau\partial_t\right)\meanemf_i
=  \alpha_0 \meanB_i - \eta_{\rm t0}\mu \meanJ_i \,,
\label{dEMFdt}
\EN
which closes the mean-field induction equation \eqref{KH03}.
In this section, we ask how useful such an approach is to model
the dynamo action of flow II
qualitatively and perhaps even quantitatively.

\subsection{A model for flow~II}

If SOCA were applicable to flow II,
\Eq{ansatz} would agree with \Eq{KH33} for $\hat\alpha$ and $\etath$.
However, to explain dynamo action we have
to go beyond SOCA, so \eqref{ansatz} can only be regarded as an approximation.
The differential equations for
$\meanB_x$ and $\meanB_y$ decouple,
which is most easily formulated by employing the mean vector potential $\meanAA$,
with $\meanBB=\nab\times\meanAA$:
\EQ
\left(1-\ell^2\partial_z^2+\tau\partial_t\right)\meanemf_i
=\pm a_0 \partial_z \meanA_i + b_0\partial_z^2 \meanA_i
\label{dEMFdtA}
\EN
which has to be solved along with
\EQ
\partial_t\meanA_i=\meanemf_i+\eta\partial_z^2\meanA_i\,,
\label{dAdt}
\EN
$i=1,2$.
In \Eq{dEMFdtA}, the upper and lower signs apply respectively to $i=1$ and 2.
If SOCA were valid, we would have
\EQ
\begin{alignedat}{2}
a_0&=\Rm v_0/4,\quad
b_0&&=\Rm v_0/8k_0,\\
\ell&=1/\sqrt{2}k_0,\quad
\tau&&=\Rm/2v_0k_0,
\end{alignedat} \label{norm}
\EN
which implies that $\ell^2/\tau=\eta$.
In the following, our corresponding non-SOCA results will sometimes
be normalized by these values.

\begin{figure}\begin{center}
\includegraphics[width=\columnwidth]{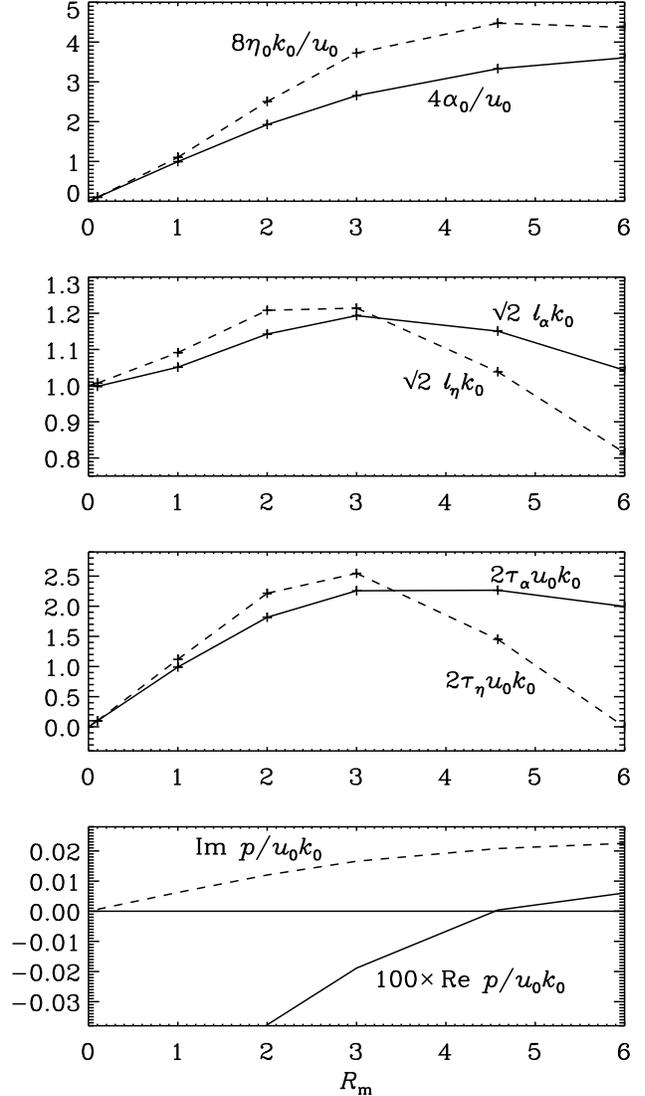}
\end{center}\caption[]{
$\Rm$ dependence of $\alpha_0$, $\etatz$, $\ell_\sigma$, and $\tau_\sigma$
for $\sigma=\alpha$ or $\eta$, flow II.
In the last panel, we plot the resulting growth rate
$p$ as obtained from \Eq{disprel} for $k/k_0=0.025$.
}\label{presults2}\end{figure}

When taking the ansatzes \eqref{ansatz} for valid, but allowing now
$\ell$ and $\tau$ to be different for
$\hat\alpha$ and $\etath$, all parameters
can be obtained from the TFM--identified dependencies
$\hat\alpha(k,\omega)$ and  $\etath(k,\omega)$
via the following recipes
\begin{align}
\alpha_0&=\lim_{k\to0}\,\mbox{Re}\,\hat\alpha(k,0), \nonumber\\
\ell_\alpha&=\frac{1}{k}\left[{\mbox{Re}\big(1/\hat\alpha(k,0)\big)
\over\mbox{Re}\big(1/\hat\alpha(0,0)\big)}-1\right]^{1/2}\;\;
\mbox{for any fixed $k\ne0$},\\
\tau_\alpha&=-\mathop{\lim_{\omega\to0}}_{k\to0}\left[{1\over\omega}
{\mbox{Im}\big(1/\hat\alpha(k,\omega)\big)\over
\mbox{Re}\big(1/\hat\alpha(k,\omega)\big)}\right], \nonumber
\end{align}
and analogously for $\etatz $, $\ell_\eta$, and $\tau_\eta$.
The results are listed in \Tab{TResults} and plotted in
\Fig{presults2} in dependence on $\Rm$,
along with the resulting growth rate, which can be obtained
by inserting \eqref{ansatz} with $\tau_\alpha=\tau_\eta=\tau$ and  $\ell_\alpha=\ell_\eta=\ell$ into \eqref{KH53} and solving for $p$:
\begin{align}
p_\pm=&\frac{\tau^{-1}+(\eta_\emf +\eta)k^2}{2}  \label{disprel} \\
&\times\left\{-1\pm\left[
1-4{\eta_\emf\eta k^4+\tau^{-1}\left(\mp\ii\alpha_0 k+\etaTz k^2\right)
\over\big(\tau^{-1}+(\eta_\emf +\eta)k^2\big)^2}
\right]^{1/2}\right\}. \nonumber
\end{align}
Here we have set $\etaTz=\etatz+\eta$ and $\eta_\emf=\ell^2/\tau$.
We have scaled the coefficients
with their respective SOCA values at $\Rm=1$,
see Eqs.~\eqref{norm}.

\begin{table}\caption{
$\Rm$ dependence of $\alpha_0$, $\etatz$, $\ell$ and $\tau$
for flow II.
}\vspace{12pt}\centerline{\begin{tabular}{r|cccccr}
$\Rm$ &  $\!\!\alpha_0/v_0\!\!$ & $\!\!\etatz k_0/v_0\!\!$ &
$\ell_\alpha k_0$ & $\!\!\tau_\alpha v_0k_0\!\!$ &
$\ell_\eta k_0$ & $\!\!\tau_\eta v_0k_0$ \\
\hline
$ 1.00$ & $ 0.249$ & $ 0.138$ & $ 0.743$ & $ 0.495$ & $ 0.772$ & $ 0.560$ \\
$ 2.00$ & $ 0.481$ & $ 0.313$ & $ 0.808$ & $ 0.908$ & $ 0.854$ & $ 1.108$ \\
$ 3.00$ & $ 0.664$ & $ 0.465$ & $ 0.844$ & $ 1.128$ & $ 0.859$ & $ 1.273$ \\
$ 4.58$ & $ 0.832$ & $ 0.559$ & $ 0.814$ & $ 1.133$ & $ 0.734$ & $ 0.725$ \\
$ 6.00$ & $ 0.901$ & $ 0.546$ & $ 0.737$ & $ 0.998$ & $ 0.576$ & $-0.001$ \\
\label{TResults}\end{tabular}}\end{table}

\begin{figure}\begin{center}
\includegraphics[width=\columnwidth]{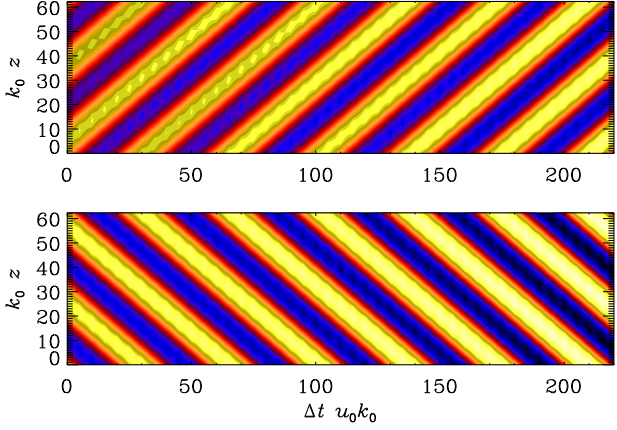}
\end{center}\caption[]{
$\meanB_x$ (top) and $\meanB_y$ (bottom) in a $zt$ diagram as obtained from
the mean-field model \eqref{dEMFdtA}, \eqref{dAdt}
for flow II with $\Rm=6$ and a $z$ extent of $20 \pi/k_0$,
using the parameters of \Tab{TResults}
and random initial conditions.
The (for the chosen $z$ extent) fastest growing mode has $k/k_0=0.2$,
the growth rate is $0.00184\,v_0k_0$ and the frequency is $0.17\,v_0k_0$.
}\label{pbutter_128b}\end{figure}

\begin{figure}\begin{center}
\hspace*{-6.5mm}\includegraphics[width=1.1\columnwidth]{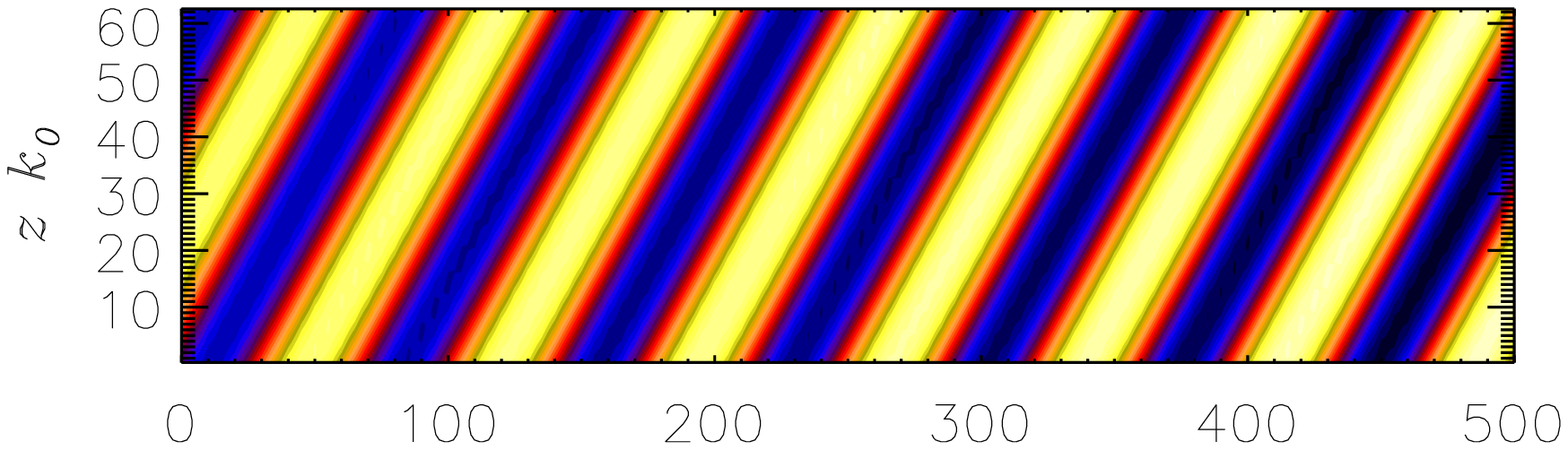} \\[-7mm]
\hspace*{-6.5mm}\includegraphics[width=1.1\columnwidth]{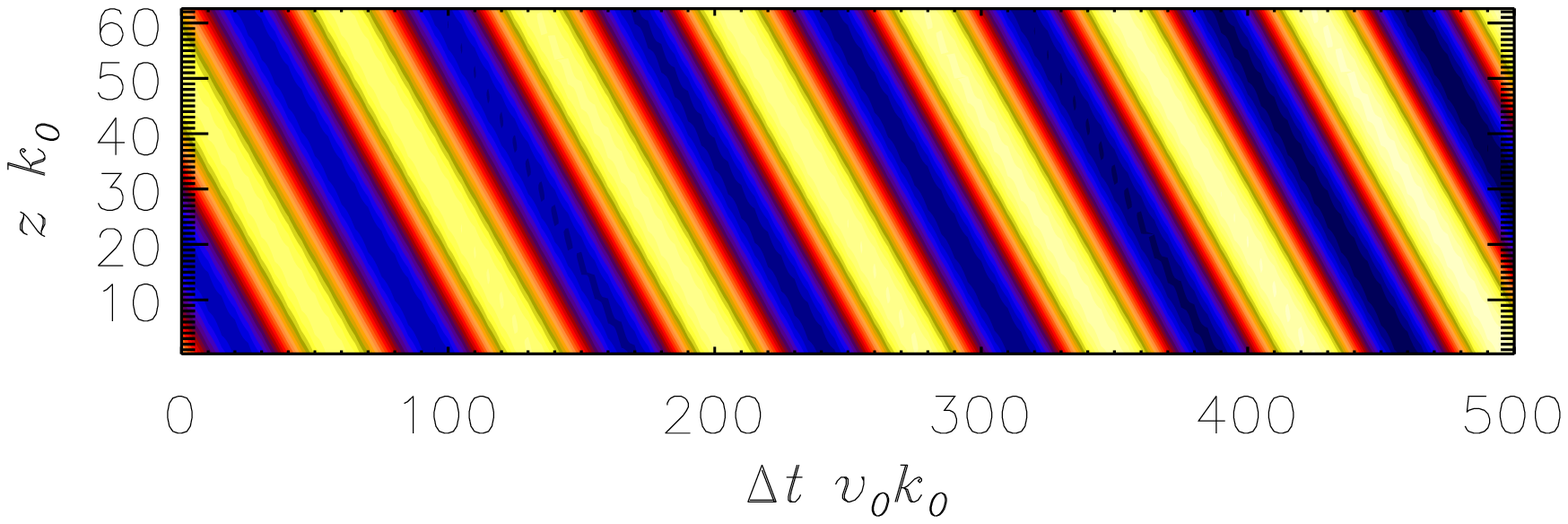}
\end{center}
\vspace{-4mm}
\caption[]{
$\meanB_x$ (top) and $\meanB_y$ (bottom)
in a $zt$ diagram as obtained from
the mean-field model \eqref{dEMFdtA}, \eqref{dAdt} for flow II with
$\Rm=5$
and a $z$ extent of $20 \pi/k_0$,
using interpolated parameters
and random initial conditions.
The (for the chosen $z$ extent) fastest growing mode has $k=0.1 k_0$ while
growth rate and oscillation frequency are $0.000545 \,v_0k_0$ and $0.0861  \,v_0k_0$, respectively.
For the definition of $\Delta t$ see \Fig{pbutter_64x192_251_R6_ran}.
}\label{pbutter_128c}\end{figure}

\begin{figure}\begin{center}
\hspace*{-6.5mm}\includegraphics[width=1.1\columnwidth]{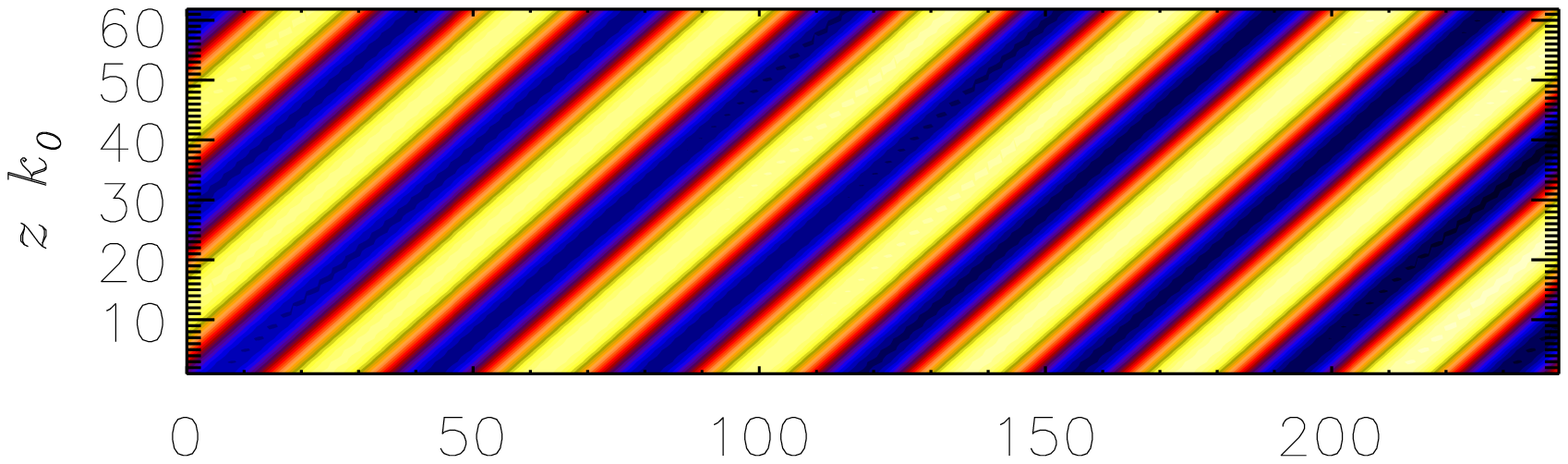} \\[-7mm]
\hspace*{-6.5mm}\includegraphics[width=1.1\columnwidth]{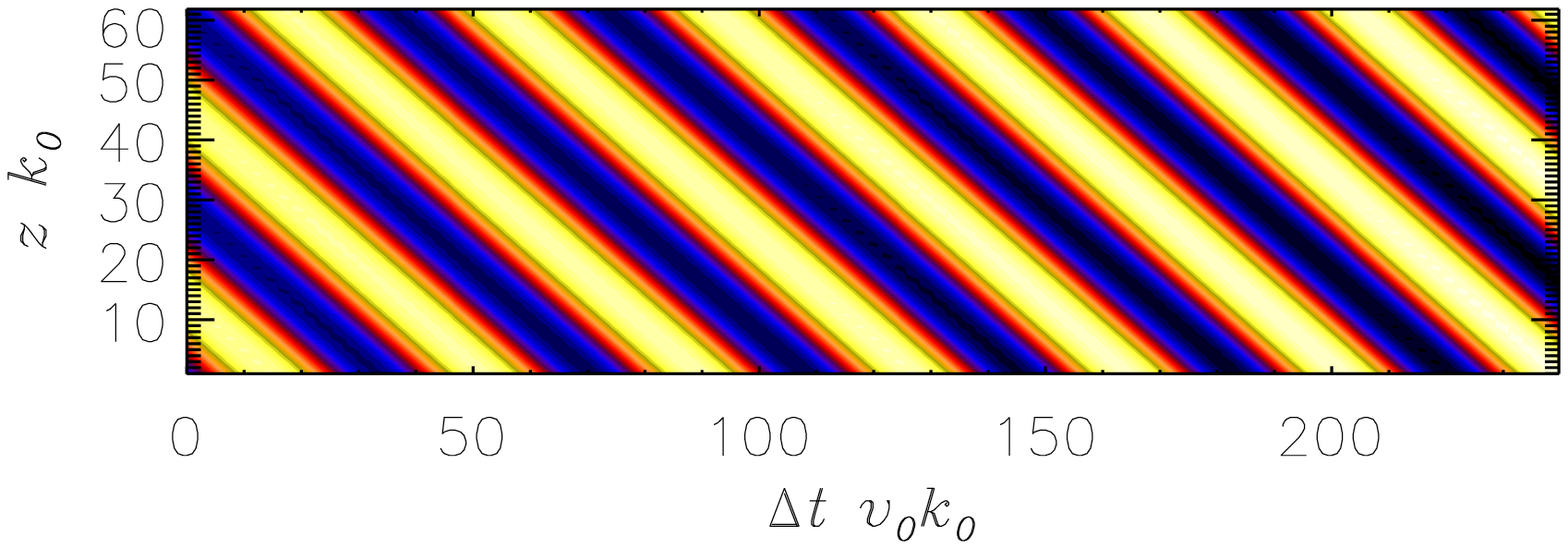}
\end{center}

\vspace{-4mm}
\caption[]{
$\meanB_x$ and $\meanB_y$ in a $zt$ diagram as obtained from
DNS for flow II with $\Rm=5$ and a $z$ extent of $20 \pi/k_0$, using random initial conditions.
The (for the chosen $z$ extent) fastest growing mode has $k=0.2 k_0$ while
growth rate and frequency are $0.00074\,v_0k_0$
and  $0.168\,v_0 k_0$, respectively.
The corresponding values for the mode with $k=0.1 k_0$ are $0.00044  \,v_0k_0$ and  $ 0.0854  \,v_0k_0$.
}\label{pbutter_64aDNS}\end{figure}

We now employ the resulting mean-field coefficients to solve the underlying
system of mean-field equations.
For that purpose we use again the {\sc Pencil Code}, which comes with a
corresponding mean-field module to solve \Eqs{dEMFdtA}{dAdt}.
In \Fig{pbutter_128b} we show the resulting mean-field components,
$\meanB_x$ and $\meanB_y$, in a $zt$ diagram.
Note that there are propagating waves traveling in opposite directions
for $\meanB_x$ and $\meanB_y$.
This result agrees qualitatively with that of the DNS
(\Fig{pbutter_64x192_251_R6_ran}), but the growth rate is too small
($0.00184\,v_0k_0$ instead of $0.00408\,v_0k_0$)
and also the (for the chosen $z$ extent)
most unstable wavenumber is too low ($0.2\,k_0$ instead of $0.3\,k_0$).

We choose $\Rm=5$ and, interpolating in Table\ \ref{TResults},
$\alpha_0=0.87 v_0$, $\etatz=0.562 v_0/k_0$,
$\tau=\tau_\alpha=1.1/v_0 k_0$, and $\ell=\ell_\alpha=0.799/k_0$,
where the latter choices are somewhat arbitrary given that $\tau_\alpha\ne\tau_\eta$ and $\ell_\alpha\ne\ell_\eta$.
In \Fig{pbutter_128c} we show the resulting $zt$ diagram
for $\meanB_x$ and $\meanB_y$.
Again, for both components there are propagating waves,
but traveling in opposite directions.
The result agrees qualitatively with that of the DNS when restricting it to $k=0.1 k_0$
(\Fig{pbutter_64aDNS}), but the growth rate is somewhat too big,
$5.45\times10^{-4} \,v_0k_0$ vs. $4.4\times10^{-4} \,v_0k_0$ from DNS,
whereas the oscillation frequencies match well:
$0.0861  \,v_0k_0$ vs. $ 0.0854  \,v_0k_0$ from DNS.
However, for a $z$ extent of $20\pi/k_0$,
the fastest growing mode has twice the wavenumber, $k=0.2 k_0$,
along with $\lambda=7.4\times10^{-4}\,v_0k_0$ and $\omega=0.168\,v_0 k_0$,
which are also almost twice as large.
The corresponding prediction from a mean-field simulation is
$\lambda=2.9\times10^{-4}\,v_0k_0$, $\omega=0.168\,v_0k_0$.

\section{Conclusions}

The present work has demonstrated a qualitatively new mean-field dynamo behavior that works chiefly through
a memory effect.
Without it, the examples of flows II and III studied in this paper would yield just decaying oscillatory solutions.
Remarkable is also the fact that flow II has zero kinetic helicity pointwise.
This, together with the fact that the two relevant components of the
mean magnetic field evolve completely independently of each other
(one can be zero, for example), might lead one to the suggestion that
the mean-field dynamo behavior of flow II could be
due to negative eddy diffusivity.
However, unlike flow IV, where the real part of the total diffusivity
(sum of turbulent and microphysical magnetic diffusivities) is indeed negative
when dynamo action occurs,
it is for flows II and III not only positive,
but turbulent and microphysical contributions have
the same order of magnitude.
The sum of these two positive contributions has to be overcome by
additional inductive effects to produce growing solutions.
These inductive effects come from the symmetric off-diagonal
components of the $\alpha$ tensor combined with the memory effect.

It is unclear how generic this qualitatively new mean-field dynamo
behavior is.
Off-diagonal components of the ${\bm\alpha}$ tensor are commonly found
in inhomogeneous turbulence, but then they are usually antisymmetric and
thus correspond to turbulent pumping.
Not much attention has yet been paid toward
symmetric off-diagonal contributions of ${\bm\alpha}$.
However, we do know that in convection such contributions do exist
in all the cases with shear; see Figs.~9 and 12 of \cite{KKB09}.
Dynamos owing to a combination of memory effect and otherwise non-generative or even diffusive effects
might not be restricted to off-diagonal components of ${\bm\alpha}$.
It is more generally connected with oscillatory behavior of a system combined with the memory effect.
Two other examples have been considered in the work of \cite{RB12},
where oscillatory solutions of both an inhomogeneous $\alpha^2$ dynamo
and a homogeneous $\alpha\Omega$ dynamo have produced significantly
lower critical dynamo numbers
in comparison with the model without memory effect.

The present work has highlighted the importance of using the
test-field method to diagnose the nature of large-scale dynamos.
Even without taking the memory effect into account, i.e., if the
test fields were assumed constant in time, it would have delivered
the information about the unusual occurrence of symmetric off-diagonal
components of ${\bm\alpha}$ in flow II
and of a $\gamma$ effect in flow III.

Finally, let us emphasize the usefulness of taking spatio-temporal
nonlocality even to lowest order into account.
Technically, this is straightforward by replacing the usual equation
of the mean electromotive force by a corresponding evolution equation.
This automatically ensures that the response to changes in the mean
magnetic field is causal and does not propagate with a speed faster
than the rms velocity of the turbulence, as was demonstrated by
\cite{BKM04} in connection with turbulent passive scalar diffusion.
It also guarantees that there is no mean-field response
to structures varying on small length scales that could otherwise
be artificially amplified; see a corresponding discussion in \cite{CGB11}.

Although for flow II the lowest-order nonlocal representation
employed in our mean-field calculations breaks down for relatively small
values of $\Rm$, there are reasons to believe that this is a peculiarity of the prescribed laminar flows.
In this respect, turbulent flows tend to be better behaved, as has
been demonstrated on several other occasions in comparison with the
Galloway-Proctor flow \citep{CHT06}, where complicated, non-asymptotic $\Rm$
dependences of $\alpha$ and $\gamma$
occur that are not found for turbulent flows \citep[cf.][]{SBS08,RB09}.

\section*{Acknowledgements}

Financial support from the Scientific \& Technological Research Council
of Turkey (T\"UB\.ITAK) and the European Research Council under the AstroDyn
Research Project 227952, the Swedish Research Council under the grants
621-2011-5076 and 2012-5797, as well as the Research Council of Norway
under the FRINATEK grant 231444 are gratefully acknowledged.
The computations have been carried out at the National Supercomputer
Centre in Ume{\aa} and at the Center for Parallel Computers at the
Research Project 227952, the Swedish Research Council under the grants
Royal Institute of Technology in Sweden and the
Nordic High Performance Computing Center in Iceland.



\end{document}